\documentclass[aip, jcp, showpacs, superscriptaddress, preprint ,11pt, longbibliography]{revtex4-1}
\usepackage{braket}	
\usepackage{mhchem}	
\usepackage{siunitx}	
\usepackage{amsmath}	

\usepackage{graphicx}	
\usepackage{multirow}	
\usepackage{wrapfig}	
\usepackage{dcolumn}	

\usepackage{verbatim}	
\usepackage[normalem]{ulem}	
\usepackage{lineno}	

\usepackage{xcolor}	
\usepackage{url}	

\newcommand{\wn}{cm$^{-1}\!\!$ }		
\newcommand{\um}{\si{\micro\metre}}			
\newcommand{\tnu}{\tilde{\nu}}
\newcommand{\tnuz}{\tilde{\nu}_0}
\newcommand{\SnuN}{S_{\tilde{\nu}}^N}

\newcommand{\E}[1]{$\times 10^{#1}$}		

\newcommand{\ortho}{\textit{ortho}}
\newcommand{\para}{\textit{para}}
\newcommand{\rmref}{\mathrm{ref}}
\newcommand{\thisband}{$\nu_{1}+\nu_{3}$ }

\begin{document}

\title{Re-evaluation of ortho-para-dependence of self pressure-broadening in the \thisband band of acetylene}

\author{Eisen\ C.\ Gross}\email{eisen.gross@stonybrook.edu}
\author{Kimberly\ A.\ Tsang}\email{ktsang@estee.com}\thanks{Now at: Estee Lauder Inc., Melville, NY 11747, USA}
\author{Trevor\ J.\ Sears}\email{trevor.sears@stonybrook.edu}
\affiliation{Chemistry Department, Stony Brook University, Stony Brook, NY 11794-3400, USA}

\begin{abstract}
  Optical frequency comb-referenced measurements of self pressure-broadened line profiles of the R(8) to R(13) lines in the \thisband combination band of acetylene near 1.52 \um\, are reported.  The analysis of the data found no evidence for a previously reported [Iwakuni et al. \textit{Phys. Rev. Lett.} \textbf{117}, 143902(5) 2016] systematic alternation in self pressure-broadened line widths with the nuclear spin state of the molecule. The present work brought out the need for the use of an accurate line profile model and a careful accounting for weak background absorptions due to hot band and lower abundance isotopomer lines.  The data were adequately fit using the quadratic speed-dependent Voigt profile model, neglecting the small speed-dependent shift.  Parameters describing the most probable and speed-dependent pressure-broadening, most probable shift, and the line strength were determined for each line.  Detailed modeling of the results of Iwakuni et al. showed that their neglect of collisional narrowing due to the speed-dependent broadening term, combined with the strongly absorbing data recorded and analyzed in transmission mode were the reasons for their results.
\end{abstract}

\date{\today}

\maketitle

\section{Introduction}
\label{Sec:Intro}

The absorption spectra of molecules in a gaseous sample are broadened by a combination of the Doppler effect due to the Maxwell-Boltzmann distribution of velocities, and the lifetimes of the levels involved in the spectroscopic transition.  For spectra in the infrared region, the natural lifetimes are normally very long and coherence lifetimes are limited by intermolecular collisions that alter the molecular velocity, the orientation of the molecule, or cause the molecule to transition to another quantum state entirely. As sample pressures are increased, so are the collision rates, and the widths of the Doppler-limited lines observed at low pressures are found to broaden and the line shape function changes.  Understanding the variation of the shapes and widths of spectral lines as a function of pressure and composition is critical for remote sensing and other analytical applications and much effort has been expended in the measurement, calculation and modeling of observed line profiles.\cite{Hitran2016,Hartmann2008, Hartmann2013,Hartmann2018,Nguyen2020} \\

Given our understanding of the phenomenon, it was therefore a surprise when, in  2016, Iwakuni et al.\cite{Iwakuni2016} measured the well-studied \thisband vibrational combination band of acetylene using a new instrument based on a dual frequency-comb\cite{Udem2002,Jones2000,Diddams2001,Washburn2004,Diddams2007,Mcraven_2011} spectrometer and reported a large ($\approx10\%$) alternation in the measured self pressure-broadening coefficients with rotational quantum number.  Transitions involving acetylene molecules in the \ortho\enspace nuclear spin arrangement apparently showed a greater self pressure-broadening than \para\enspace ones.  In the ground state of acetylene, \ortho\enspace nuclear spin symmetry is associated with odd rotational quantum number ($J$) levels, while \para-symmetry is associated with even $J$ levels, hence correlating with the alternation in line widths with rotational quantum number in the spectrum. \\

The observations were explained by the following arguments.  Since the \ortho-spin arrangement is statistically favored 3:1 to the \para\enspace one, \ortho-\ortho\enspace collisions are more probable than \para-\para\enspace ones, so that efficient resonant collisional energy transfer processes are more probable among \ortho\enspace molecules than \para\enspace ones.  If the collisions that cause broadening are dominated by resonant energy transfer ones, then $\Delta J=\pm2$ rotational energy transfer between molecules with common nuclear spin symmetry will be more efficient than other inelastic interactions. Having a light molecule where the rotational intervals are large compared to translational  energies helps to see the effect because the  $|\Delta J|=2$ resonant collisions become more probable than inelastic ones for large rotational energy spacings. The general idea was first described by Anderson in 1949.\cite{Anderson1949}  It is well documented in gaseous \ce{H2}, from rotational Raman spectra,\cite{Keijser1974,VanDenHout1988,Rahn1991} but \ce{H2} is a special case due to its huge rotational energy level spacings.  There have been reports of analogous effects in \ce{HCl}\cite{Gray1971,Rich1971,Fabre1972} where the rotational energy spacings are also large, resulting in a peak in the pressure-broadening coefficient at low$-J$, and an oscillatory behavior in broadening at high$-J$.  In \ce{HCN}, the observed self pressure-broadening coefficient\cite{Smith1986, Swann2005, Bouanich2005} peaks at rotational levels at the maximum of the Maxwell-Boltzmann distribution, a fact explained by resonant and near resonant rotational energy transfer processes dominating the collisional broadening. However there had been no observation of the effect in acetylene which does not possess a dipole moment and whose quadrupole moment is small. \\

The reported effects\cite{Iwakuni2016} are comparable to or greater than the limits of the precision of data recorded using conventional Fourier Transform Infrared (FTIR) spectrometers, and so it was surprising that previous experimental measurements and analyses had missed the differences.  The report was immediately controversial and theoretical calculations and analysis by Lehmann\cite{Lehmann2017} suggested that effects of quadrupole-induced $|\Delta J|=2$ resonant energy transfer could not explain the size of the measured differences. Also,he noted that the effects of resonant collisions in the upper state of the transitions would be reversed because there the odd-$J$ levels are \para \enspace symmetry and the same arguments would imply resonant collisions with odd-$J$ ground state levels would still be preferred.\\

Separately, Hartmann and Tran\cite{Hartmann2017} suggested that the use of a Voigt profile (VP) model in the original analysis was at fault because fits of the measured transmission spectra were made, and the more strongly absorbing \ortho\enspace lines lead to systematically increased apparent widths compared to the less strongly absorbing \para\enspace lines due to the model deficiencies. The authors of the original work rebutted these arguments,\cite{Iwakuni2017} but no experimental study has yet been reported to confirm or deny the experimental results or to investigate the effects of using more accurate line profile models on precise experimental data. \\

In this work, we report measurements of the $R(8)$ to $R(13)$ rotational lines in the \thisband band of pure acetylene gas in question.  These lines showed the strongest \ortho-\para\enspace pressure-broadening differences in the original report.\cite{Iwakuni2016}  The data were all recorded at (measured) ambient temperatures (295 to 300K) and over a pressure range from 0.14  to 7 kPa (1 to 52 Torr).  Measurements were made with an extended cavity diode laser (ECDL) with the frequency component phase-locked to a mode of an optical frequency comb (OFC) resulting in absolute frequencies and long-term relative optical frequency stability good to a few parts in 10$^{11}$.  The pressure-dependent data for a given rotation-vibration line were analyzed in a multispectrum fit to extract line profile parameters.  Several line profile models, derived from the Hartmann-Tran profile,\cite{Ngo2013, Tennyson2014_IUPAC} were considered for use in the analysis, and the quadratic speed-dependent Voigt profile (QSDVP) model neglecting the speed-dependent shift ($\delta_{2}$) was chosen based on its ability to reproduce the data using a minimal number of parameters.  See also Atkins and Hodges\cite{Adkins2019} for a discussion of line profile choices for the analysis of experimental data.  We found it necessary to account for the presence of weak underlying hot band and lower abundance isotopomer lines that affect the observed baselines and can distort line profiles of the spectra, but our analysis of the data showed no evidence for a systematic \ortho-\para\enspace variation in pressure-broadening.  Extensive modeling of synthetic data shows the most likely explanation for the original observations is that suggested by Hartmann and Tran.\cite{Hartmann2017}  The neglect of collisional narrowing in the Voigt line profile model used in the transmission representation and the large peak absorbances in the original work caused a systematic over-estimate of widths for the stronger absorbing \ortho\enspace lines even at pressures where collisional narrowing is small. \\

\section{Experimental Details}
\label{Sec:Experiment}
\begin{figure}[h]
 \centering
 \includegraphics[width=12cm]{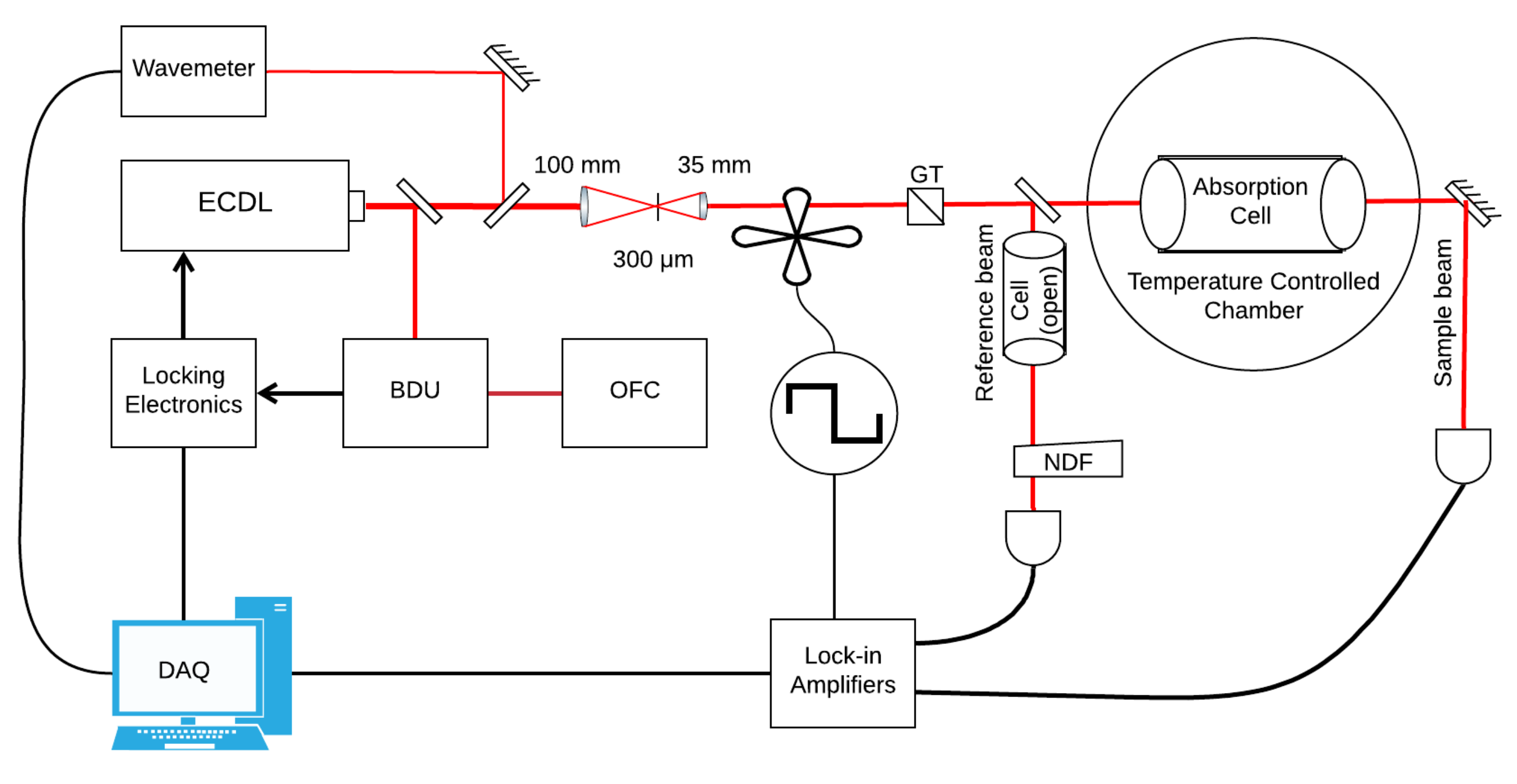}
  \caption{\label{img:experiment}\small{Diagram of experimental layout. ECDL: extended cavity diode laser, OFC: optical frequency comb, BDU: beat detection unit, DAQ: data acquisition program, GT: Glan-Taylor prism. NDF: variable neutral density filter.  The reference and sample beams are labeled in the figure.  The cell on the reference beam was open to air, and partially compensated for the optical power loses in the sample beam.}}
\end{figure}
A tunable ECDL (Sacher Lasertechnik TEC 500) centered at 1550 nm was driven by a precision current supply (ILX Lightwave LDC-3900). The output beam was passed through an optical isolator (Isowave, Inc.) and split into two.  The first beam containing 50\% of the power was coupled into an optical fiber and directed to the beat detection unit (BDU) associated with an OFC (Menlo Systems FC-1500, with a repetition rate of 250 MHz).  The second 50\% of power was split again with 4\% of it coupled into a optical fiber to a wavemeter (Bristol Instruments model 621B) for wavelength measurement.  The diameter of the remaining beam was scaled down by one-third to 0.70 mm diameter using a telescope and the spatial mode was cleaned up with a pinhole before passing through an optical chopper (Scitec, 350CD) operated at $\approx$1650 Hz.  Finally, the beam was split again using a 50:50 non-polarizing beamsplitter, creating two beam paths for sample and reference detectors. A Glan-Taylor prism was used before this final split to reduce the optical power and maintain a linear response of the detectors that are described in detail below. Typical continuous optical power at each of the detectors was estimated to be 0.07 mW under the conditions of the measurements. \\

The absorption cell was 1.085 cm long machined from solid copper with wedged, indium sealed, \ce{CaF2} windows that was used previously.\cite{Cich2013, McRaven2011}  It was held inside a metal vacuum chamber sealed with a pair of wedged silica windows.  The vacuum chamber incorporated a temperature-controlled mounting assembly giving a convenient way to monitor the cell temperature, while also providing temperature stabilization and gas leak prevention.  To compensate for the two pairs of windows in the sample beam, the reference beam was directed through a second cell (open to the air) and a variable neutral density filter to maintain equal optical power levels at the two detectors in the absence of any absorbing sample. \\

Finally, the sample and reference beams were focused onto a matched pair of detectors to provide sample transmission, $I_t$, and reference, $I_0$  signals.  The two detectors were based on Hamamatsu InGaAs photodiodes (model G8605-11), with a negative-biased preamplifying circuit using an AD6202 op-amp. Their output signals were analyzed by two lock-in amplifiers (SRS, model SR830) controlled by LabView code over a GPIB bus, with the lock-in reference frequency provided by the chopper controller.  The lock-in amplifiers use a 24-bit Motorola DSP56001 chip clocked at 30 MHz, corresponding to a digitized Nyquist frequency near 0.06 $\mu$s.  The time constant for the lock-in amplifiers was 100 ms and the signals were averaged for one second before transferring to the data acquisition computer.  Non-absorbing signals from the detectors were typically 160 mV.   \\

The output of the OFC, referenced to a 10 MHz GPS clock signal, was overlapped with one of the beams from the ECDL inside the BDU, and the power levels were balanced using polarized optics.  The signal from the BDU was used to generate an error-signal which was applied to the cavity-length piezoelectric translator, PZT, of the ECDL, establishing a stable frequency output.  The exact beat frequency and its Allan variance was measured using a frequency counter (SRS SR620), and recorded. The beat frequency was measured to have a full width-half maximum of less than 1 MHz, and the Allan variance was typically 4 kHz over periods of several minutes.\\

The natural abundance acetylene gas sample was purified by cryo-distillation at -97$^\circ$ C. Infrared spectra of the sample before purification showed traces of impurities including acetone and organic sulfides.  Following purification all impurity infrared signatures were reduced by at least a factor of 50.  The sample of purified acetylene gas was loaded into the absorption cell from a gas handling manifold to the desired pressure, and the cell was closed to the gas manifold and the pressure recorded.  The manifold was evacuated using a turbomolecular pump.  A capacitance  pressure gauge (MKS-398HD) was used for all sample pressure measurements.  Its absolute calibration was checked using a new piezoelectric transducer (MKS 902B-11010), and the maximum measurement difference was found to be 0.24\%.  For a given sample pressure, data were recorded for the lines (R(8) to R(13)) in the vibrational band by scanning the down-mixed repetition rate frequency of the OFC.  A scan of 12 kHz at 50 Hz increments result in a 2.1 GHz wide optical spectrum with a point spacing of 7.8 MHz.  For the lowest two sets of pressures, one scan was sufficient, however most data were recorded using two consecutive scans, resulting in a 4.2 GHz wide spectra.  This ensured an adequate amount of baseline was recorded for stable line profile fitting.  Once all the lines for a given pressure were recorded, the manifold was closed to the pumping, the cell re-opened to it, and a second pressure measurement made. The known relative volumes of the cell and manifold allowed any cell pressure change during the measurements to be checked and no changes outside the gauge error limits were detected during the measurements.  Experimental conditions for the recorded lines are listed in Table \ref{tab:conditions}. \\

\begin{table}[ht]
 \centering
  \caption{\label{tab:conditions}\small{Pressures used for the measurements, in kilo-pascal and torr.}}
\scriptsize
\begin{tabular}{ccc}
 \hline
  Pressure  & Pressure    & Lines \\
   kPa    &  Torr         & Measured \\
\hline
  7.0099  &  52.579	  &  all lines \\
  4.1394  &  31.048  & R(9), R(11), R(13) \\
  4.0045  &  30.036  & R(8), R(10), R(12) \\
  2.2035  &  16.528   &  all lines  \\
  1.091  &  8.184	  &  all lines \\
  0.6393 &  4.795	  &  all lines \\
  0.1384 &  1.038	  &  all lines \\
  \hline
\end{tabular}
\normalsize
\end{table} 

\section{Analysis}
\label{Sec:Analysis}

\subsection{Line Modeling}
\label{SSec:LineModel}
 Multispectrum fits of the sets of data corresponding to the same vibration-rotational transition at varying pressures (in Table \ref{tab:conditions}) were performed.  A suite of Python programs, described and available as Supplementary Information for this paper, were written to read the data files, collate files by transition, fit the data to a normalized line shape function, and present the resulting line profile parameters. To avoid complications associated with a logarithmic transformation of the data we fit the raw transmittance data.  Hartmann and Tran\cite{Hartmann2017} showed fitting with absorbance eliminates systematic problems at large absorption strengths with the VP model, and we have confirmed this, see Section \ref{sec:modeling} below.  At the lower peak absorbances and higher pressures used in this work, compared to those of Iwakuni et al.,\cite{Iwakuni2016} even fitting the present data to a VP in the transmission representation did not show the systematic \ortho-\para\enspace broadening variation they reported. \\

 Power transmittance, $t$, is usually defined as:

\begin{equation} \label{eq:tran1}
  t(\tnu ) = \frac{I_t(\tnu)}{I_0(\tnu)}
\end{equation}
where $\tnu$ is the wavenumber frequency of light, $I_t$ is the intensity of a beam transmitted through a sample gas and $I_0$ is the initial beam intensity.  According to the Beer-Lambert law, transmittance can be expressed as:

\begin{equation} \label{eq:tran2}
  t(\tnu)= e^{-\SnuN(T) \times N \times l_{\mathrm{path}} \times g(\tnu - \tnuz)}. 
\end{equation}
Here, $g(\tnu - \tnuz)$ is the area normalized line profile function for the molecular absorption centered around the line wavenumber, $\tnu_0$,  optical pathlength is $l_{\mathrm{path}}$, and $N$ is the number density of the gas, calculated by assuming the ideal gas law, such that $N = p/(k_{\mathrm{B}}T)$, with $p$ the pressure, and $k_{\mathrm{B}}$ the Boltzmann constant.  $S_{\tnu}^{N}(T)$ is the spectral line strength which is dependent on temperature, $T$.  Simeckova et al.\cite{Simeckova2006} show how $\SnuN$ is related to the Einstein A-coefficient for the transition. \\

The HITRAN\cite{Hitran2016} database reports values of $S_{HIT}\,=\,S_{\tnu}^{N}(T_{\mathrm{ref}})$ for a natural isotopic distribution of sample molecules at a reference temperature $T_{\mathrm{ref}}\, =\, 296$ K, and these can be converted to other temperatures using Eq. (\ref{eq:spectint}), from \textcite{Simeckova2006}.
\begin{equation} \label{eq:spectint}
  S_{\tnu}^{N}(T) = S_{HIT} \frac{Q_{tot}(T_{\rmref})}{Q_{tot}(T)} e^{-c_{2}\tilde{E}''(1/T - 1/T_{\rmref})} \left[ \frac{1-e^{c_{2}\tnu/T}}{1-e^{c_{2}\tnu/T_{\rmref}}} \right]
\end{equation}
where $\tilde{E}''$ is the wavenumber corresponding to the energy of the lower level of the transition, also reported in HITRAN,\cite{Hitran2016} the function $Q_{tot}(T)$ is the total internal partition sum (TIPS) that describes the statistical population of states at thermal equilibrium and which is discussed below, and $c_{2}$ is a constant derived from other constants:
\begin{equation} 
  c_{2} = \frac{hc}{k_{\mathrm{B}}}.  \notag 
\end{equation}
    Here $h$ is the Planck constant and $c$ is the speed of light.   Numerical evaluation of the direct partition function sum can require a heavy computational load, however the result can be well approximated with a polynomial function, such as the one given in Eq. (\ref{eq:partition}), \cite{Gamache2000} over a limited temperature range. 

\begin{equation} \label{eq:partition}
  Q_{tot}(T) \approx -8.3088 + 1.4484 \left(T/\si{\kelvin}\right) - 2.5946\times10^{-3} \left(T/\si{\kelvin}\right)^{2} + 8.4612\times10^{-6}\left( T/\si{\kelvin}\right)^{3}
\end{equation}

  We evaluated Eq. (\ref{eq:partition}) against other published empirical results\cite{Amyay2011} which showed a difference of less than 0.14\% at temperatures near to 300 K.  We also coded a direct numerical summation using molecular constants from Robert et al.\cite{Robert2007}  The results agreed with Eq. (\ref{eq:partition}) to better than 0.04\% at the temperatures of interest in the present work, and also with the results of a more recent tabulation.\cite{Gamache2017}  Details of these comparisons have been included in the Supplementary Information for this paper.\\

Equation (\ref{eq:tran2}) shows how the transmittance data, $t(\tnu)$ is related to the normalized molecular line shape, $g(\tnu-\tnuz)$.  Since the IUPAC (International Union of Pure and Applied Chemistry) recommended\cite{Tennyson2014_IUPAC} the adoption of the line shape proposed by Ngo et al.\cite{Ngo2013} in 2014, and named it the Hartmann-Tran Profile (HTP), it has increasingly become the lineshape model of choice in high resolution spectroscopy.  Fitting to the full set of parameters of the HTP requires extensive and very high signal-to-noise ratio (SNR) data, however the HTP has the advantage of being able to be reduced to several simpler line profile models\cite{Ngo2013} by nulling specific parameters. The Gaussian, Doppler-broadening, contribution\cite{Townes1975} for all profile models was calculated for the given temperature, and was not varied in this work.  \\ 

  The commonly used VP is obtained by eliminating all but the two most important contributions to the collisionally perturbed profile: the pressure-broadening, $\gamma$, and shift, $\delta$, coefficients.  Adkins et al. \cite{Adkins2019} analyzed the effects of the smaller HTP parameters showing they are sensitive to different noise characteristics in the data. Therefore, we experimented using different profile functions with various combinations of adjustable parameters.  The  QSDVP gave the most reliable fit of the present data (see below) with fewest parameters varying.  It includes two contributions to both the pressure-broadening and pressure-shift terms: a contribution dependent on the most probable speed, $\gamma_0$ and $\delta_0$ respectively, and speed-dependent line-broadening, $\gamma_2$, and shift, $\delta_2$, corrections.  However, the higher-order pressure-dependent shift coefficient was not well determined in preliminary fits to the present data, which do not extend to sufficiently high pressure to determine it reliably, and $\delta_2$ was fixed at zero for the data fitting discussed below.  \\

\subsection{Corrections for underlying lines}
\begin{figure}[h]
 \centering
 \includegraphics[width=12cm]{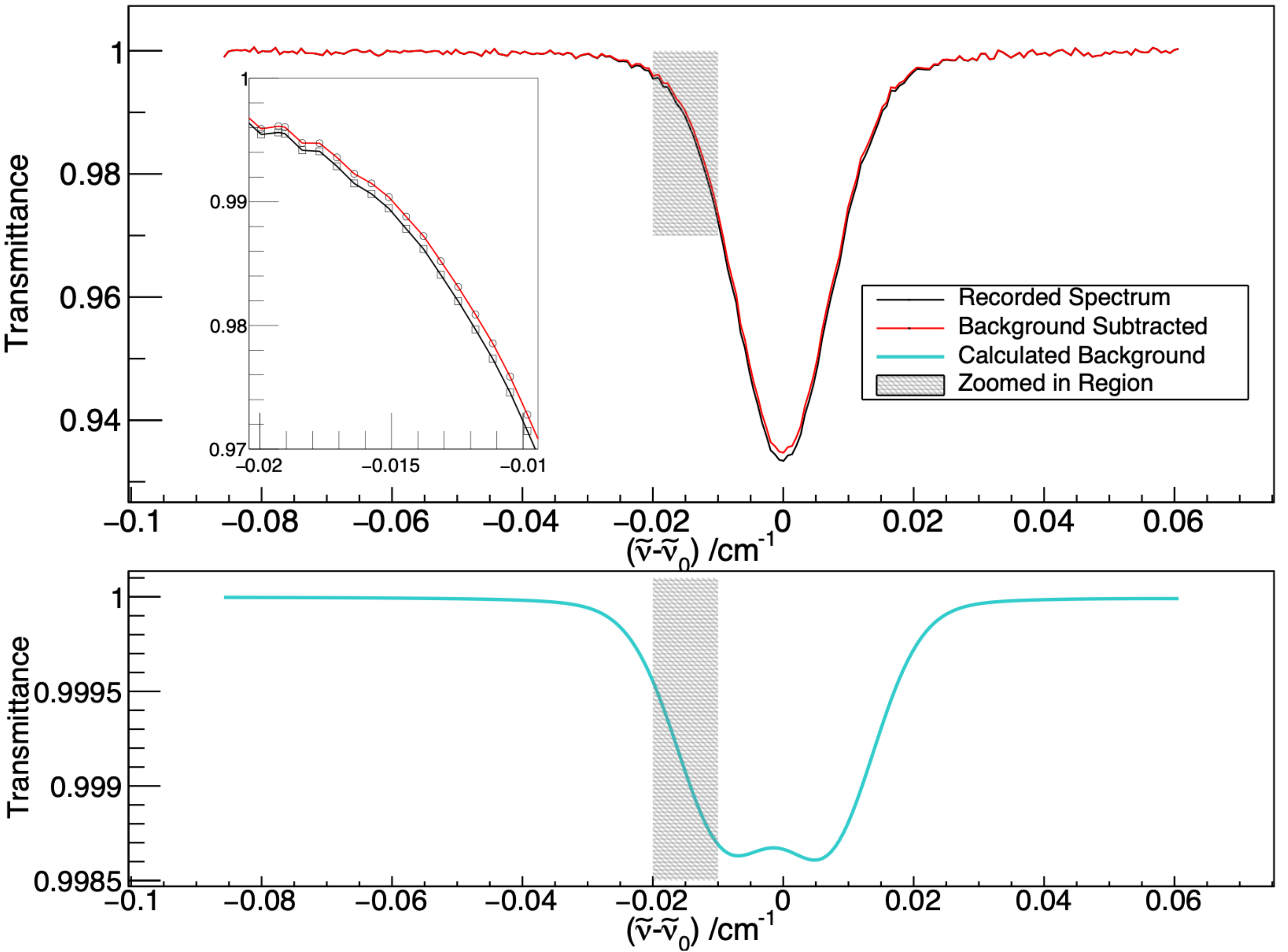}
  \caption{\label{img:hotband}\small{An example of the influence of background transitions near R(10), (1.091 kPa, 295.94 K).  The calculated background line spectrum on the lower panel, asymmetrically broadens the line of interest, as is seen in the zoomed region (highlighted).  The two background transitions are assigned\cite{Hitran2016} to  P$_e$(17) of the transition from the ground state to the level 110(11)$^{0\,+}_{u}$ at 6580.6349 \wn, and the R$_f$(24) line of the hot band from \textit{v}$_4$ = 1 to the level 101(10)$^1_g$ at 6580.6511 \wn.  The vibrational quantum numbers here are $v_1,v_2,v_3(v_4,v_5)$ with the superscript the total vibrational angular momentum, $l\,=\,l_4+l_5$, the $g$ or $u$ symmetry and $+$ or $-$ for sigma levels.}} 
\end{figure}

The effects of weaker hot band and low abundance isotope lines lying close to the transitions of interest are not negligible.  Background spectra including these weaker lines for specific regions of pressure and temperature were generated for each line by combining Voigt profiles using parameters from HITRAN\cite{Hitran2016}, and included in the Supplementary Information for this paper.  The frequency points from the actual data were used as the frequency variable to create the appropriate background function.  This resulted in a one-to-one correspondence of the data points between the observed and background spectra making background subtraction much simpler.  Accounting for background lines was a vital part of the data analysis since many weaker transitions lie near or underneath the transition of interest, and have absorptions of $\approx 1\%$ of the main line.  Figure \ref{img:hotband} shows the results for the R(10) transition where the neglect of the background transitions causes a strong distortion in the profile.  \\

  Several weak background lines in the data were not found in the HITRAN\cite{Hitran2016} list.  They affected R(8) and R(12).  Adjustments were made to empirically correct for these observed features by adding a VP feature. All the background lines included in the analysis are tabulated in Supplementary Information for this paper.

\subsection{Baseline and amplitude fitting}
As in our previous work,\cite{Forthomme2015} small instrumental baseline deviations were accommodated into the profile fitting function.  The baseline of each transition at each pressure had to be individually fitted.  An initial set of parameters for a given transition together with hot band and weaker underlying features described above, were used to estimate a line profile and an apparent baseline function derived by subtracting it from the data. This was then fit to a second order polynomial to provide three parameters to fit the baseline, and the full model function is given as Eq. (\ref{eq:fullFit}). 
\begin{align} \label{eq:fullFit}
  e^{- a_{0}\, \SnuN(T) \, N \, l_{\mathrm{path}}\, g(\tnu - \tnu_{0})} &= \left( \frac{I_t(\tnu)}{I_0(\tnu)}\right)_{\mathrm{obs}}\!\! - \big[b_{0} + b_{1}(\tnu - \tnu_{0}) + b_{2}(\tnu - \tnu_{0})^{2} \big] \\ \notag &-\sum_b^{\mathrm{b.g.\,lines}}\left( e^{-\SnuN(T) \, N \, l_{\mathrm{path}}\, g_{b}(\tnu - \tnu_{0})} \right)_b,
\end{align}
where the $I_{t} / I_{0}$ term represents the recorded data, the polynomial function represents the baseline subtraction, the summed term accounts for the weak underlying lines discussed in the previous section, and the exponential term to the left is the expected line transmittance.  We include a scalar factor, $a_0$, to accommodate small deviations from the expected line strength, $\SnuN(T)$, see below. \\

 The baseline fitting process was repeated after new line profile parameters were determined from the multispectrum fitting for a given transition until convergence.  Since the baseline was instrumental, rather than associated with the spectra, the baseline was subtracted linearly from the data rather than including it in the exponential term,\cite{Forthomme2015} providing a more stable fit.  The line strengths, $S_{\tnu}^N(T)$, from Eq. \ref{eq:spectint} were found to closely reproduce the observed intensities, but to accommodate the small variations, a dimensionless scalar multiplier, $a_0$, expected to be close to unity, was introduced as a correction factor.  The lowest-pressure scans, where the fixed Gaussian contribution to the line profiles dominates, were used in preliminary fitting to determine the amplitude corrections by floating the multiplier $a_0$.  These multiplier values were then reevaluated over cycles of the multispectrum fit for each line until the final fit when all parameters were varied to yield the results in Table \ref{tab:fitresults}. 

\section{Results}
\label{Sec:Results}

\begin{figure}[h]
 \centering
   \includegraphics[width=12cm]{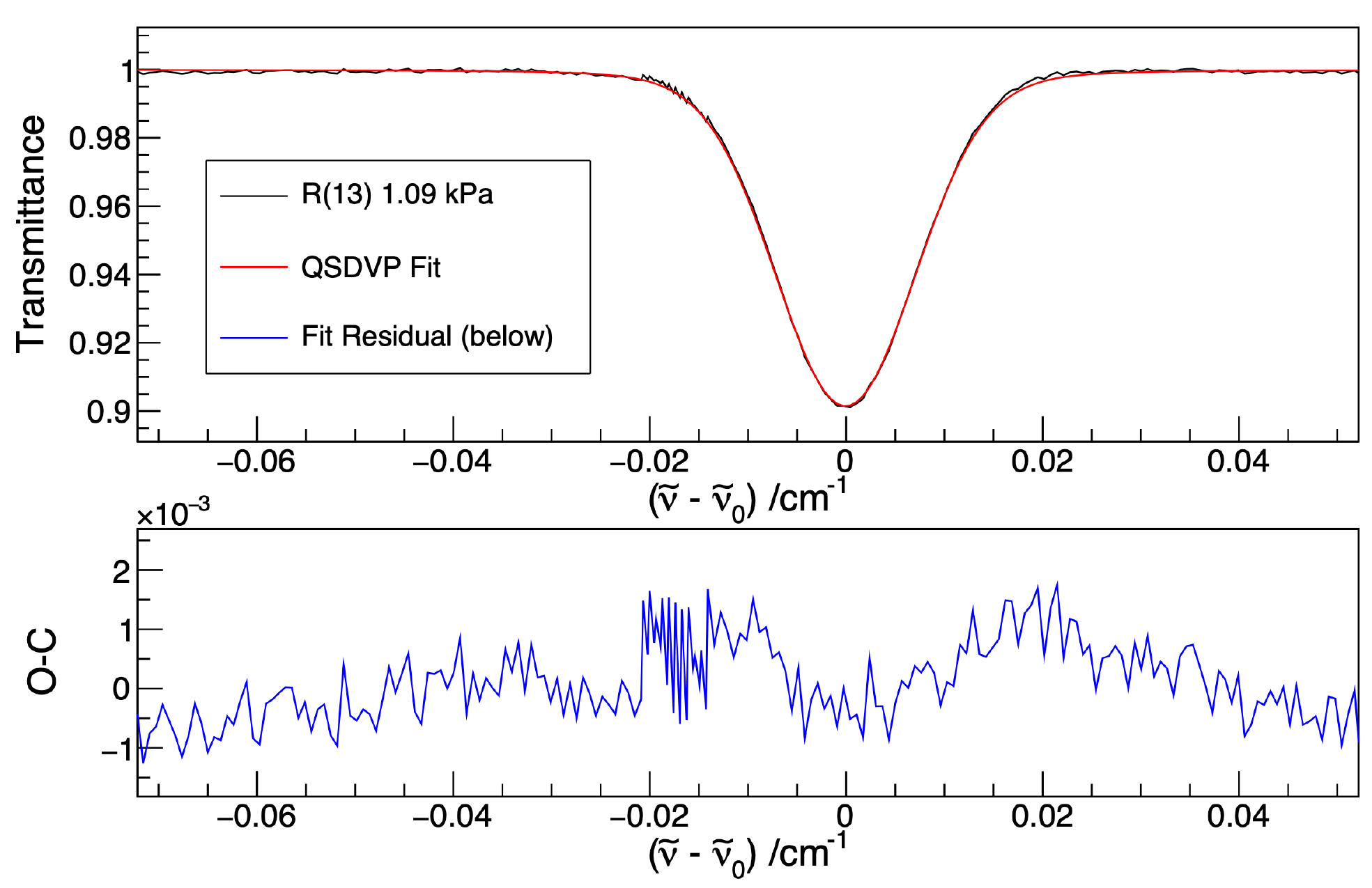}
  \caption{\label{img:fit}\small{Observed and calculated spectra of R(13) at 1.091 kPa (8.184 Torr) and 294.4 K.  The calculated lines shape was generated with the QSDVP function as detailed in the text, with the observed - calculated residual in the lower panel.  The additional points around $\tnu-\tnuz=-0.02$ \wn\, are due to overlap between successive scans.}}
\end{figure}

  \subsection{Line profile fitting}
  As a typical example, Fig. \ref{img:fit} illustrates the results of fitting the data for R(13) at 1.091 kPa (8.184 Torr).   Table \ref{tab:fitresults} presents the fitted parameters to the reduced QSDVP discussed above.  The line strength multiplier values found are all slightly larger than, but within $\approx$2\% of, unity and consistent with that seen in our previous work.\cite{Forthomme2015}  The results also agree with those from another recent measurement of the line strengths in this band.\cite{Okubo2017}   The differences between the newly measured line strengths and those derived from Eq. \ref{eq:spectint}  are factors of 10 larger than can be accounted for by errors in the TIPS polynomial approximation, discussed in section \ref{Sec:Analysis}.  \\

\begin{table}[ht]
 \centering
  \caption{\label{tab:fitresults}\small{Results of the line profile modeling$^a$.}}
\scriptsize
\begin{tabular}{ccccc}
 \hline
  Line Label & {$\gamma_{0}$ } & {$\delta_{0}$ } & {$\gamma_{2}^b$ } &  $a_0$\\
\hline
R(8)  & 0.15517 (40)$^c$ &	-6.87 (23)  \E{-3}   & 2.261 (77) \E{-2}     & 1.0236 (10)		\\
R(9)  & 0.15055 (16) &	-7.122 (90) \E{-3}   & 2.121 (29) \E{-2}     & 1.01796 (39)             \\
R(10) & 0.14669 (39) &	-9.14 (22)  \E{-3}   & 2.059 (72) \E{-2}     & 1.0101 (11)              \\
R(11) & 0.14543 (15) &	-10.03 (87) \E{-3}  & 1.643 (27) \E{-2}     & 1.00553 (42)             \\
R(12) & 0.14453 (41) &	-10.30 (24) \E{-3}  & 1.832 (78) \E{-2}     & 1.0159 (12)              \\
R(13) & 0.14175 (17) &	-10.57 (10) \E{-3}  & 1.947 (31) \E{-2}     & 1.01898 (46)             \\
\hline
\multicolumn{5}{l}{a. All parameters have units of cm$^{-1}$/atm except $a_0$, which is dimensionless.}\\
\multicolumn{5}{l}{b. The quadratic shift, $\delta_{2}$ could not be determined, see text for details.} \\
\multicolumn{5}{l}{c. The numbers in parenthesis are one$-\sigma$ uncertainties in units of the last } \\
\multicolumn{5}{l}{ quoted significant figure.  Systematic uncertainties caused by underlying lines add a} \\
\multicolumn{5}{l}{ factor of two or more to these standard deviations, see section \ref{SSec:ErrorE}.} \\
\end{tabular}
\normalsize
\end{table} 

\begin{table}[ht]
 \centering
  \caption{\label{tab:simulParms}\small{Parameters used for line simulations, and initial fit parameters.$^a$}}
\scriptsize
\begin{tabular}{lcccccc}
 \hline
  Line	& Wavenumber\,$^b$ &  $S_{HIT} \,^{c}$	& $\tilde{E}''\,^d$	& $\gamma_0\,^e $ & $\delta_0\,^{f}$ \\
\hline
  R(7)	&  6574.361355 &	1.303\E{-20}	& 65.887      & 0.162   & -0.0071  \\ 
  R(8)	&  6576.481658 & 	4.465\E{-21}	& 84.710 		& 0.158 & -0.0077   \\
  R(9)	&  6578.575894 & 	1.340\E{-20}	& 105.885		& 0.154 & -0.0082   \\
  R(10) &  6580.644073 & 	4.380\E{-21}	& 129.411		& 0.150 & -0.0088   \\
  R(11) &  6582.686025 & 	1.264\E{-20}  & 155.289		& 0.147 & -0.0093   \\
  R(12) &  6584.701858 & 	3.977\E{-21}  & 183.517		& 0.144 & -0.0098   \\
  R(13) &  6586.691493 & 	1.107\E{-20}  & 214.096		& 0.141 & -0.0103   \\
  R(14) &  6588.654889 & 	3.369\E{-21}  & 247.024		& 0.138 & -0.0108   \\
 \hline
\multicolumn{7}{l}{a. The coefficients $\gamma_2 =$ 0.0212 cm$^{-1}$atm$^{-1}$ and $\delta_2 =$ -0.00132 } \\
\multicolumn{7}{l}{ cm$^{-1}$atm$^{-1}$ were assumed for all lines.\cite{Forthomme2015}} \\
\multicolumn{7}{l}{b. In \wn \, from reference \cite{Madej2006}} \\
\multicolumn{7}{l}{c. Spectral line strength\cite{Hitran2016} in units of \wn/(molecule $\cdot$ cm$^{-2}$)} \\
\multicolumn{7}{l}{   at $T_{\mathrm{ref}}$ = 296 K.} \\
\multicolumn{7}{l}{d. Lower state energy\cite{Hitran2016} in \wn .}\\
\multicolumn{7}{l}{e. Pressure broadening coefficient\cite{Lehmann2017,Kusaba2001,Jacquemart2003} in cm$^{-1}$atm$^{-1}$ } \\
\multicolumn{7}{l}{f. Pressure shift coefficient from Iwakuni et al.\cite{Iwakuni2016} in cm$^{-1}$atm$^{-1}$}. \\
\end{tabular}
\normalsize
\end{table} 

\begin{figure}[h]
 \centering
   \includegraphics[width=12cm]{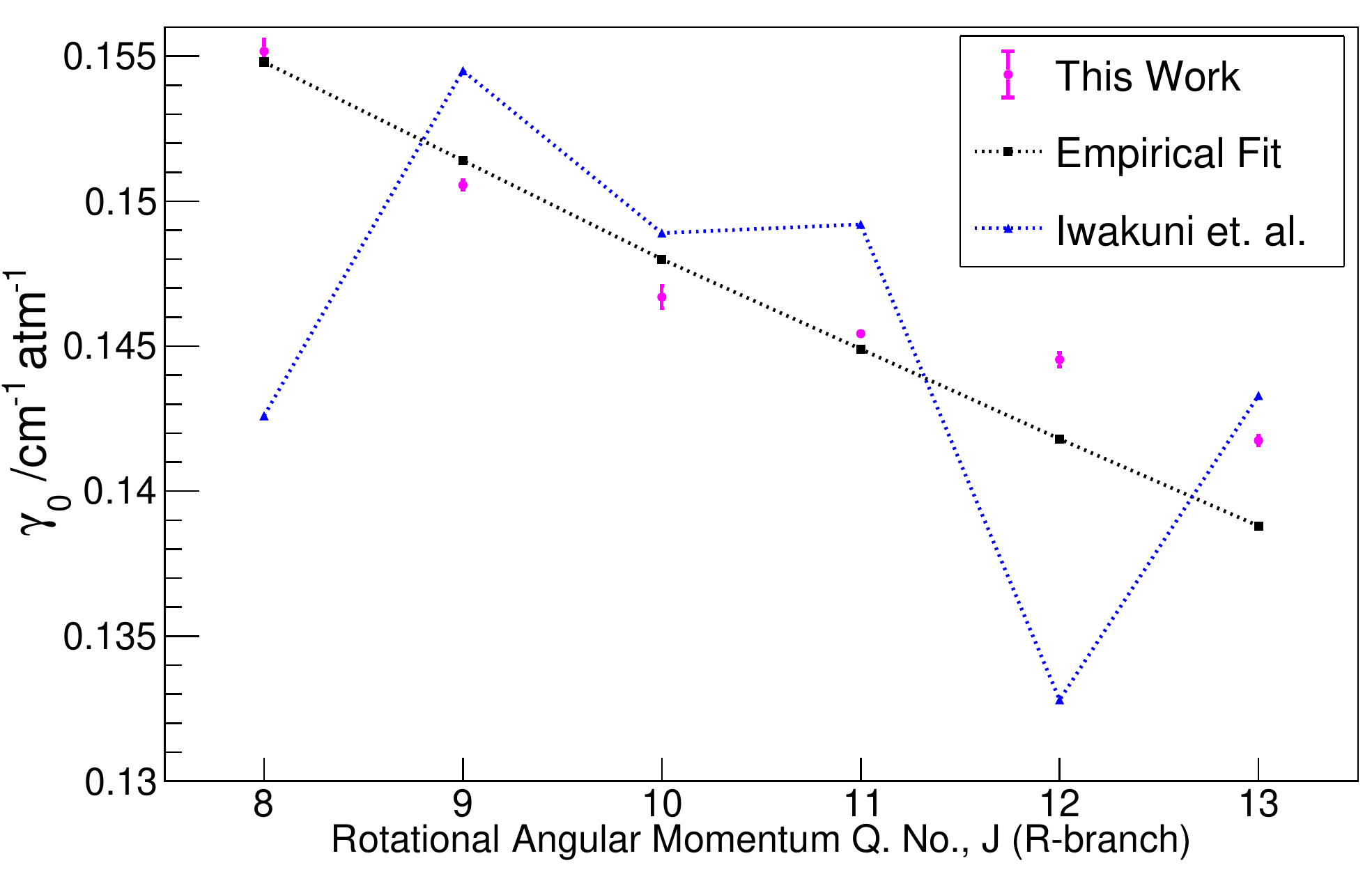}
  \caption{\label{img:compare}\small{The comparison between the QSDVP most probable speed-dependent ($\gamma_0$), and VP ($\gamma$) self pressure-broadening parameters, with 1-$\sigma$ standard deviations for the lines R(8) to R(13) in units of cm$^{-1}$atm$^{-1}$.  Empirical Fit from \cite{Lehmann2017,Kusaba2001,Jacquemart2003}, and Iwakuni et al.\cite{Iwakuni2016}}}
\end{figure}

There is a decreasing trend in the most probable speed self-broadening coefficient $\gamma_{0}$ with rotational quantum number, $J$, which is expected as rotational energy intervals increase.  The results show reasonable agreement with published empirical fits\cite{Kusaba2001,Jacquemart2003} of multiple published results to smooth functional forms, and the HITRAN database, as is illustrated in Fig. \ref{img:compare}.  Importantly, there is no evidence for the kind of systematic reduction in the average self-broadening parameter for \para\enspace lines compared to \ortho\enspace ones reported by Iwakuni et al. and also shown in this figure.  This disagreement is most likely due to the reasons outlined by Hartmann and Tran,\cite{Hartmann2017} and is investigated more thoroughly in subsection \ref{sec:modeling} below.  

\subsection{Estimates of Systematic Uncertainties}\label{SSec:ErrorE}
It frequently happens\cite{Forthomme2015} that the statistical parameter uncertainties derived from multispectrum fits of high precision data such as performed here, are underestimates of the actual values if systematic uncertainties are not taken into account. The baseline noise in the data is well represented by a Gaussian distribution with amplitude of approximately $\pm 0.0005$ in transmittance units, mostly due to source and detector noise.  Systematic measurement uncertainties in pressure and temperature or inaccurate accounting for underlying lines may have greater effects on the least-squares fitted parameter standard deviations.  \\

Systematic uncertainties in the multispectrum fitted parameters propagated by inaccurate pressure measurements were estimated in a Monte Carlo type analysis described previously.\cite{Forthomme2015}  A synthetic dataset for a single rotational transition (R(9) was selected) was generated using the parameters in Table \ref{tab:fitresults}.  The pressure in the dataset was randomly modified following a Gaussian distribution with a width, $\Delta p$,  and the multispectrum reduced QSDVP fit performed.  This was repeated 256 times to generate a distribution of parameter values whose standard deviation was determined.  The process was repeated for greater and smaller assumed $\Delta p$ values. \\  

For pressure measurement uncertainties, $\Delta p$, at the level of the specification of the gauge head, the parameter deviations were similar to the statistical standard deviations quoted in Table \ref{tab:fitresults}. As $\Delta p$ is increased, 1$\sigma$ parameter deviations increased approximately linearly.  For example, a 1\% increase in $\Delta p$ results in a 0.1\% increase, $\Delta\gamma_0$, in the $\gamma_0$ uncertainty, $\Delta \delta_0$ increases by 0.01\%, $\Delta \gamma_2$ increases by 0.05\% and $\Delta a_0$ increases by 0.1\%.  Contributions to systematic uncertainties due to any reasonable assumed inaccuracy in temperature measurements are negligible. \\

The largest contributions to the uncertainties derive from inaccuracies in the treatment of weak underlying lines which was investigated by adding a background line with a line intensity fixed at a (small) fraction of the main line at various frequency offsets from the center of the main line.  Modeling showed that the maximum strength of a background feature that would be unrecognized was at the level of approximately 0.001 times the $\SnuN(T)$ of the main line.  For R(9) this led to a variation in $\gamma_0$, $\Delta\gamma_0$, of $\pm0.0003$ cm$^{-1}$ atm$^{-1}$, \textit{i.e.} about twice the statistical uncertainty in Table \ref{tab:fitresults}.  Results for $\Delta \gamma_2$, $\Delta \delta_0$, and $\Delta a_0$ were $\pm0.02, \pm0.03$ and $\pm0.00005$ \wn\,atm$^{-1}$ respectively, all larger than their statistical uncertainties in Table \ref{tab:fitresults}. Similar relative increases can be expected in the parameter uncertainties for the other lines.  The VP $\gamma$ parameter assumed for the background lines is also a concern, since it is typically not well measured and values assumed by extrapolation from sparse data.  Modeling showed that a change of 0.01 cm$^{-1}$atm$^{-1}$ in the assumed broadening coefficient for some background lines caused changes to the $\gamma_0$ values given in Table \ref{tab:fitresults} outside the statistical uncertainties. \\

\subsection{Modeling VP fits to experimental data in the low transmittance regime} \label{sec:modeling}
Hartmann and Tran\cite{Hartmann2017} claimed that the VP is an unreliable line shape model when applied to transmission spectral data recorded under conditions when peak absorbances approach unity, especially at low pressures.  It is noticeable that in Fig. 1 of the Iwakuni et al. rebuttal \cite{Iwakuni2017} of this criticism which showed data for the R(9) transition, the line center is close to 100\% absorbing under nearly all their experimental conditions. To simulate the data and investigate the parameter space, we have modified the MATLAB code developed previously. \cite{Forthomme2015}  This simulates a line in a transmittance spectrum using a HTP model, then attempts to fit it to a simplified model.  Full details and links to GitLab repositories are provided in the Supplementary Information for this paper.  We conclude that the comment of Hartmann and Tran\cite{Hartmann2017} is valid.  The use of a VP to fit strongly absorbing transmittance data can lead to the systematic variation in self-broadening reported by Iwakuni et al.\cite{Iwakuni2016}  \\

\section{Conclusions and Summary}
 Rotation-vibration lines with rotational quantum numbers between 8 and 13 in the R-branch of the \thisband band of acetylene have been measured using an OFC-locked ECDL spectrometer.  The measurements did not support the alternation in size of the pressure self-broadening coefficient, reported by Iwakuni et al.\cite{Iwakuni2016,Iwakuni2017}  Detailed modeling of their results strongly suggested that the reported alternation is an artifact due to their simulation and fitting of the observed transmission data using the Voigt profile that neglects collisional narrowing in a strongly absorbing regime. \\

\section{Supplementary Information}
\label{Sec:SuppMat}
List of items included as supplementary material for this paper:
\begin{enumerate}
 \item Hot band and low abundance isotope lines included in the modeling of the experimental data.
 \item Details of the simulations used for modeling the data of Iwakuni et al \cite{Iwakuni2016, Iwakuni2017} with a VP.
 \item Matlab codes for calculating the Internal Partition Function for acetylene.
 \item Details of Python codes for data analysis and links to GitLab repositories.
\end{enumerate}

\section{Data Availability Statement}
The experimental data that supports the findings of this study are available from the corresponding author upon reasonable request.  The data analysis programs are detailed in the Supplementary Information of this paper, and available in the GitLab repository, \url{https://gitlab.com/searssbu/rotfit}.  The MATLAB codes used for modeling are also available in the GitLab repository, \url{https://gitlab.com/searssbu/htp-matlab_codes}.

\section{Acknowledgements} 
This work was supported by the U.S. Department of Energy, Office of Science, Division of Chemical Sciences, Geosciences and Biosciences within the Office of Basic Energy Sciences, under Award Number DE-SC0018950.  This research made use of the MINUIT algorithm\cite{James1975} via the iminuit\cite{iminuit2020} Python interface. We are most grateful for discussions and correspondence with Professors Sasada (Keio University, Japan), Hartmann (IPLS, Ecole Polytechnique, France) and Lehmann (University of Virginia) during the course of this work. \\

\footnotesize

\begin{thebibliography}{43}%
\makeatletter
\providecommand \@ifxundefined [1]{%
 \@ifx{#1\undefined}
}%
\providecommand \@ifnum [1]{%
 \ifnum #1\expandafter \@firstoftwo
 \else \expandafter \@secondoftwo
 \fi
}%
\providecommand \@ifx [1]{%
 \ifx #1\expandafter \@firstoftwo
 \else \expandafter \@secondoftwo
 \fi
}%
\providecommand \natexlab [1]{#1}%
\providecommand \enquote  [1]{``#1''}%
\providecommand \bibnamefont  [1]{#1}%
\providecommand \bibfnamefont [1]{#1}%
\providecommand \citenamefont [1]{#1}%
\providecommand \href@noop [0]{\@secondoftwo}%
\providecommand \href [0]{\begingroup \@sanitize@url \@href}%
\providecommand \@href[1]{\@@startlink{#1}\@@href}%
\providecommand \@@href[1]{\endgroup#1\@@endlink}%
\providecommand \@sanitize@url [0]{\catcode `\\12\catcode `\$12\catcode
  `\&12\catcode `\#12\catcode `\^12\catcode `\_12\catcode `\%12\relax}%
\providecommand \@@startlink[1]{}%
\providecommand \@@endlink[0]{}%
\providecommand \url  [0]{\begingroup\@sanitize@url \@url }%
\providecommand \@url [1]{\endgroup\@href {#1}{\urlprefix }}%
\providecommand \urlprefix  [0]{URL }%
\providecommand \Eprint [0]{\href }%
\providecommand \doibase [0]{http://dx.doi.org/}%
\providecommand \selectlanguage [0]{\@gobble}%
\providecommand \bibinfo  [0]{\@secondoftwo}%
\providecommand \bibfield  [0]{\@secondoftwo}%
\providecommand \translation [1]{[#1]}%
\providecommand \BibitemOpen [0]{}%
\providecommand \bibitemStop [0]{}%
\providecommand \bibitemNoStop [0]{.\EOS\space}%
\providecommand \EOS [0]{\spacefactor3000\relax}%
\providecommand \BibitemShut  [1]{\csname bibitem#1\endcsname}%
\let\auto@bib@innerbib\@empty
\bibitem [{\citenamefont {Gordon}\ \emph {et~al.}(2017)\citenamefont {Gordon},
  \citenamefont {Rothman}, \citenamefont {Hill}, \citenamefont {Kochanov},
  \citenamefont {Tan}, \citenamefont {Bernath}, \citenamefont {Birk},
  \citenamefont {Boudon}, \citenamefont {Campargue}, \citenamefont {Chance},
  \citenamefont {Drouin}, \citenamefont {Flaud}, \citenamefont {Gamache},
  \citenamefont {Hodges}, \citenamefont {Jacquemart}, \citenamefont
  {Perevalov}, \citenamefont {Perrin}, \citenamefont {Shine}, \citenamefont
  {Smith}, \citenamefont {Tennyson}, \citenamefont {Toon}, \citenamefont
  {Tran}, \citenamefont {Tyuterev}, \citenamefont {Barbe}, \citenamefont
  {Császár}, \citenamefont {Devi}, \citenamefont {Furtenbacher},
  \citenamefont {Harrison}, \citenamefont {Hartmann}, \citenamefont {Jolly},
  \citenamefont {Johnson}, \citenamefont {Karman}, \citenamefont {Kleiner},
  \citenamefont {Kyuberis}, \citenamefont {Loos}, \citenamefont {Lyulin},
  \citenamefont {Massie}, \citenamefont {Mikhailenko}, \citenamefont
  {Moazzen-Ahmadi}, \citenamefont {Müller}, \citenamefont {Naumenko},
  \citenamefont {Nikitin}, \citenamefont {Polyansky}, \citenamefont {Rey},
  \citenamefont {Rotger}, \citenamefont {Sharpe}, \citenamefont {Sung},
  \citenamefont {Starikova}, \citenamefont {Tashkun}, \citenamefont {Auwera},
  \citenamefont {Wagner}, \citenamefont {Wilzewski}, \citenamefont {Wcisło},
  \citenamefont {Yu},\ and\ \citenamefont {Zak}}]{Hitran2016}%
  \BibitemOpen
  \bibfield  {author} {\bibinfo {author} {\bibfnamefont {I.}~\bibnamefont
  {Gordon}}, \bibinfo {author} {\bibfnamefont {L.}~\bibnamefont {Rothman}},
  \bibinfo {author} {\bibfnamefont {C.}~\bibnamefont {Hill}}, \bibinfo {author}
  {\bibfnamefont {R.}~\bibnamefont {Kochanov}}, \bibinfo {author}
  {\bibfnamefont {Y.}~\bibnamefont {Tan}}, \bibinfo {author} {\bibfnamefont
  {P.}~\bibnamefont {Bernath}}, \bibinfo {author} {\bibfnamefont
  {M.}~\bibnamefont {Birk}}, \bibinfo {author} {\bibfnamefont {V.}~\bibnamefont
  {Boudon}}, \bibinfo {author} {\bibfnamefont {A.}~\bibnamefont {Campargue}},
  \bibinfo {author} {\bibfnamefont {K.}~\bibnamefont {Chance}}, \bibinfo
  {author} {\bibfnamefont {B.}~\bibnamefont {Drouin}}, \bibinfo {author}
  {\bibfnamefont {J.-M.}\ \bibnamefont {Flaud}}, \bibinfo {author}
  {\bibfnamefont {R.}~\bibnamefont {Gamache}}, \bibinfo {author} {\bibfnamefont
  {J.}~\bibnamefont {Hodges}}, \bibinfo {author} {\bibfnamefont
  {D.}~\bibnamefont {Jacquemart}}, \bibinfo {author} {\bibfnamefont
  {V.}~\bibnamefont {Perevalov}}, \bibinfo {author} {\bibfnamefont
  {A.}~\bibnamefont {Perrin}}, \bibinfo {author} {\bibfnamefont
  {K.}~\bibnamefont {Shine}}, \bibinfo {author} {\bibfnamefont {M.-A.}\
  \bibnamefont {Smith}}, \bibinfo {author} {\bibfnamefont {J.}~\bibnamefont
  {Tennyson}}, \bibinfo {author} {\bibfnamefont {G.}~\bibnamefont {Toon}},
  \bibinfo {author} {\bibfnamefont {H.}~\bibnamefont {Tran}}, \bibinfo {author}
  {\bibfnamefont {V.}~\bibnamefont {Tyuterev}}, \bibinfo {author}
  {\bibfnamefont {A.}~\bibnamefont {Barbe}}, \bibinfo {author} {\bibfnamefont
  {A.}~\bibnamefont {Császár}}, \bibinfo {author} {\bibfnamefont
  {V.}~\bibnamefont {Devi}}, \bibinfo {author} {\bibfnamefont {T.}~\bibnamefont
  {Furtenbacher}}, \bibinfo {author} {\bibfnamefont {J.}~\bibnamefont
  {Harrison}}, \bibinfo {author} {\bibfnamefont {J.-M.}\ \bibnamefont
  {Hartmann}}, \bibinfo {author} {\bibfnamefont {A.}~\bibnamefont {Jolly}},
  \bibinfo {author} {\bibfnamefont {T.}~\bibnamefont {Johnson}}, \bibinfo
  {author} {\bibfnamefont {T.}~\bibnamefont {Karman}}, \bibinfo {author}
  {\bibfnamefont {I.}~\bibnamefont {Kleiner}}, \bibinfo {author} {\bibfnamefont
  {A.}~\bibnamefont {Kyuberis}}, \bibinfo {author} {\bibfnamefont
  {J.}~\bibnamefont {Loos}}, \bibinfo {author} {\bibfnamefont {O.}~\bibnamefont
  {Lyulin}}, \bibinfo {author} {\bibfnamefont {S.}~\bibnamefont {Massie}},
  \bibinfo {author} {\bibfnamefont {S.}~\bibnamefont {Mikhailenko}}, \bibinfo
  {author} {\bibfnamefont {N.}~\bibnamefont {Moazzen-Ahmadi}}, \bibinfo
  {author} {\bibfnamefont {H.}~\bibnamefont {Müller}}, \bibinfo {author}
  {\bibfnamefont {O.}~\bibnamefont {Naumenko}}, \bibinfo {author}
  {\bibfnamefont {A.}~\bibnamefont {Nikitin}}, \bibinfo {author} {\bibfnamefont
  {O.}~\bibnamefont {Polyansky}}, \bibinfo {author} {\bibfnamefont
  {M.}~\bibnamefont {Rey}}, \bibinfo {author} {\bibfnamefont {M.}~\bibnamefont
  {Rotger}}, \bibinfo {author} {\bibfnamefont {S.}~\bibnamefont {Sharpe}},
  \bibinfo {author} {\bibfnamefont {K.}~\bibnamefont {Sung}}, \bibinfo {author}
  {\bibfnamefont {E.}~\bibnamefont {Starikova}}, \bibinfo {author}
  {\bibfnamefont {S.}~\bibnamefont {Tashkun}}, \bibinfo {author} {\bibfnamefont
  {J.~V.}\ \bibnamefont {Auwera}}, \bibinfo {author} {\bibfnamefont
  {G.}~\bibnamefont {Wagner}}, \bibinfo {author} {\bibfnamefont
  {J.}~\bibnamefont {Wilzewski}}, \bibinfo {author} {\bibfnamefont
  {P.}~\bibnamefont {Wcisło}}, \bibinfo {author} {\bibfnamefont
  {S.}~\bibnamefont {Yu}}, \ and\ \bibinfo {author} {\bibfnamefont
  {E.}~\bibnamefont {Zak}},\ }\bibfield  {title} {\enquote {\bibinfo {title}
  {The {HITRAN2016} molecular spectroscopic database},}\ }\href {\doibase
  https://doi.org/10.1016/j.jqsrt.2017.06.038} {\bibfield  {journal} {\bibinfo
  {journal} {J. Quant. Spectros. Radiat. Transfer}\ }\textbf {\bibinfo {volume}
  {203}},\ \bibinfo {pages} {3 -- 69} (\bibinfo {year} {2017})}\BibitemShut
  {NoStop}%
\bibitem [{\citenamefont {Hartmann}, \citenamefont {Boulet},\ and\
  \citenamefont {Robert}(2008)}]{Hartmann2008}%
  \BibitemOpen
  \bibfield  {author} {\bibinfo {author} {\bibfnamefont {J.}~\bibnamefont
  {Hartmann}}, \bibinfo {author} {\bibfnamefont {C.}~\bibnamefont {Boulet}}, \
  and\ \bibinfo {author} {\bibfnamefont {D.}~\bibnamefont {Robert}},\ }\href
  {\doibase 10.1016/B978-0-444-52017-3.X0001-5} {\emph {\bibinfo {title}
  {Collisional Effects on Molecular Spectra}}}\ (\bibinfo  {publisher}
  {Elsevier},\ \bibinfo {year} {2008})\BibitemShut {NoStop}%
\bibitem [{\citenamefont {Hartmann}\ \emph {et~al.}(2013)\citenamefont
  {Hartmann}, \citenamefont {Tran}, \citenamefont {Ngo}, \citenamefont
  {Landsheere}, \citenamefont {Chelin}, \citenamefont {Lu}, \citenamefont
  {Liu}, \citenamefont {Hu}, \citenamefont {Gianfrani}, \citenamefont {Casa},
  \citenamefont {Castrillo}, \citenamefont {{{Lep\`ere}}}, \citenamefont
  {{{Deli\`ere}}}, \citenamefont {Dhyne},\ and\ \citenamefont
  {Fissiaux}}]{Hartmann2013}%
  \BibitemOpen
  \bibfield  {author} {\bibinfo {author} {\bibfnamefont {J.-M.}\ \bibnamefont
  {Hartmann}}, \bibinfo {author} {\bibfnamefont {H.}~\bibnamefont {Tran}},
  \bibinfo {author} {\bibfnamefont {N.~H.}\ \bibnamefont {Ngo}}, \bibinfo
  {author} {\bibfnamefont {X.}~\bibnamefont {Landsheere}}, \bibinfo {author}
  {\bibfnamefont {P.}~\bibnamefont {Chelin}}, \bibinfo {author} {\bibfnamefont
  {Y.}~\bibnamefont {Lu}}, \bibinfo {author} {\bibfnamefont {A.-W.}\
  \bibnamefont {Liu}}, \bibinfo {author} {\bibfnamefont {S.-M.}\ \bibnamefont
  {Hu}}, \bibinfo {author} {\bibfnamefont {L.}~\bibnamefont {Gianfrani}},
  \bibinfo {author} {\bibfnamefont {G.}~\bibnamefont {Casa}}, \bibinfo {author}
  {\bibfnamefont {A.}~\bibnamefont {Castrillo}}, \bibinfo {author}
  {\bibfnamefont {M.}~\bibnamefont {{{Lep\`ere}}}}, \bibinfo {author}
  {\bibfnamefont {Q.}~\bibnamefont {{{Deli\`ere}}}}, \bibinfo {author}
  {\bibfnamefont {M.}~\bibnamefont {Dhyne}}, \ and\ \bibinfo {author}
  {\bibfnamefont {L.}~\bibnamefont {Fissiaux}},\ }\bibfield  {title} {\enquote
  {\bibinfo {title} {Ab-initio calculations of the spectral shapes of
  {{CO}}$_2$ isolated lines including non-{V}oigt effects and comparisons with
  experiments},}\ }\href@noop {} {\bibfield  {journal} {\bibinfo  {journal}
  {Phys. Rev. A}\ }\textbf {\bibinfo {volume} {87}},\ \bibinfo {pages} {013403}
  (\bibinfo {year} {2013})}\BibitemShut {NoStop}%
\bibitem [{\citenamefont {Hartmann}\ \emph {et~al.}(2018)\citenamefont
  {Hartmann}, \citenamefont {Tran}, \citenamefont {Armante}, \citenamefont
  {Boulet}, \citenamefont {Campargue}, \citenamefont {Forget}, \citenamefont
  {Gianfrani}, \citenamefont {Gordon}, \citenamefont {Guerlet}, \citenamefont
  {Gustafsson}, \citenamefont {Hodges}, \citenamefont {Kassi}, \citenamefont
  {Lisak}, \citenamefont {Thibault},\ and\ \citenamefont
  {Toon}}]{Hartmann2018}%
  \BibitemOpen
  \bibfield  {author} {\bibinfo {author} {\bibfnamefont {J.-M.}\ \bibnamefont
  {Hartmann}}, \bibinfo {author} {\bibfnamefont {H.}~\bibnamefont {Tran}},
  \bibinfo {author} {\bibfnamefont {R.}~\bibnamefont {Armante}}, \bibinfo
  {author} {\bibfnamefont {C.}~\bibnamefont {Boulet}}, \bibinfo {author}
  {\bibfnamefont {A.}~\bibnamefont {Campargue}}, \bibinfo {author}
  {\bibfnamefont {F.}~\bibnamefont {Forget}}, \bibinfo {author} {\bibfnamefont
  {L.}~\bibnamefont {Gianfrani}}, \bibinfo {author} {\bibfnamefont
  {I.}~\bibnamefont {Gordon}}, \bibinfo {author} {\bibfnamefont
  {S.}~\bibnamefont {Guerlet}}, \bibinfo {author} {\bibfnamefont
  {M.}~\bibnamefont {Gustafsson}}, \bibinfo {author} {\bibfnamefont {J.~T.}\
  \bibnamefont {Hodges}}, \bibinfo {author} {\bibfnamefont {S.}~\bibnamefont
  {Kassi}}, \bibinfo {author} {\bibfnamefont {D.}~\bibnamefont {Lisak}},
  \bibinfo {author} {\bibfnamefont {F.}~\bibnamefont {Thibault}}, \ and\
  \bibinfo {author} {\bibfnamefont {G.~C.}\ \bibnamefont {Toon}},\ }\bibfield
  {title} {\enquote {\bibinfo {title} {Recent advances in collisional effects
  on spectra of molecular gases and their practical consequences},}\ }\href
  {\doibase https://doi.org/10.1016/j.jqsrt.2018.03.016} {\bibfield  {journal}
  {\bibinfo  {journal} {J. Quant. Spectros. Radiat. Transfer}\ }\textbf
  {\bibinfo {volume} {213}},\ \bibinfo {pages} {178 -- 227} (\bibinfo {year}
  {2018})}\BibitemShut {NoStop}%
\bibitem [{\citenamefont {Nguyen}, \citenamefont {Ngo},\ and\ \citenamefont
  {Tran}(2020)}]{Nguyen2020}%
  \BibitemOpen
  \bibfield  {author} {\bibinfo {author} {\bibfnamefont {H.~T.}\ \bibnamefont
  {Nguyen}}, \bibinfo {author} {\bibfnamefont {N.~H.}\ \bibnamefont {Ngo}}, \
  and\ \bibinfo {author} {\bibfnamefont {H.}~\bibnamefont {Tran}},\ }\bibfield
  {title} {\enquote {\bibinfo {title} {Line-shape parameters and their
  temperature dependences predicted from molecular dynamics simulations for
  {{O}}$_2$ and air-broadened {{CO}}$_2$ lines},}\ }\href {\doibase
  https://doi.org/10.1016/j.jqsrt.2019.106729} {\bibfield  {journal} {\bibinfo
  {journal} {J. Quant. Spectros. Radiat. Transfer}\ }\textbf {\bibinfo {volume}
  {242}},\ \bibinfo {pages} {106729} (\bibinfo {year} {2020})}\BibitemShut
  {NoStop}%
\bibitem [{\citenamefont {Iwakuni}\ \emph {et~al.}(2016)\citenamefont
  {Iwakuni}, \citenamefont {Okubo}, \citenamefont {Yamada}, \citenamefont
  {Onae}, \citenamefont {Hong},\ and\ \citenamefont {Sasada}}]{Iwakuni2016}%
  \BibitemOpen
  \bibfield  {author} {\bibinfo {author} {\bibfnamefont {K.}~\bibnamefont
  {Iwakuni}}, \bibinfo {author} {\bibfnamefont {S.}~\bibnamefont {Okubo}},
  \bibinfo {author} {\bibfnamefont {K.~M.}\ \bibnamefont {Yamada}}, \bibinfo
  {author} {\bibfnamefont {A.}~\bibnamefont {Onae}}, \bibinfo {author}
  {\bibfnamefont {F.-L.}\ \bibnamefont {Hong}}, \ and\ \bibinfo {author}
  {\bibfnamefont {H.}~\bibnamefont {Sasada}},\ }\bibfield  {title} {\enquote
  {\bibinfo {title} {Ortho-para dependent pressure effects observed in the near
  infrared band of acetylene by dual-comb spectroscopy},}\ }\href {\doibase
  10.1063/1.1726843} {\bibfield  {journal} {\bibinfo  {journal} {Phys. Rev.
  Lett.}\ }\textbf {\bibinfo {volume} {117}},\ \bibinfo {pages} {143902(5)}
  (\bibinfo {year} {2016})}\BibitemShut {NoStop}%
\bibitem [{\citenamefont {Udem}, \citenamefont {Holzwarth},\ and\ \citenamefont
  {H\"{a}nsch}(2002)}]{Udem2002}%
  \BibitemOpen
  \bibfield  {author} {\bibinfo {author} {\bibfnamefont {T.}~\bibnamefont
  {Udem}}, \bibinfo {author} {\bibfnamefont {R.}~\bibnamefont {Holzwarth}}, \
  and\ \bibinfo {author} {\bibfnamefont {T.~W.}\ \bibnamefont {H\"{a}nsch}},\
  }\bibfield  {title} {\enquote {\bibinfo {title} {Optical frequency
  metrology},}\ }\href {\doibase 10.1038/416233a} {\bibfield  {journal}
  {\bibinfo  {journal} {Nature}\ }\textbf {\bibinfo {volume} {416}},\ \bibinfo
  {pages} {233--237} (\bibinfo {year} {2002})}\BibitemShut {NoStop}%
\bibitem [{\citenamefont {Jones}(2000)}]{Jones2000}%
  \BibitemOpen
  \bibfield  {author} {\bibinfo {author} {\bibfnamefont {D.~J.}\ \bibnamefont
  {Jones}},\ }\bibfield  {title} {\enquote {\bibinfo {title}
  {Carrier-{Envelope} {Phase} {Control} of {Femtosecond} {Mode}-{Locked}
  {Lasers} and {Direct} {Optical} {Frequency} {Synthesis}},}\ }\href {\doibase
  10.1126/science.288.5466.635} {\bibfield  {journal} {\bibinfo  {journal}
  {Science}\ }\textbf {\bibinfo {volume} {288}},\ \bibinfo {pages} {635--639}
  (\bibinfo {year} {2000})}\BibitemShut {NoStop}%
\bibitem [{\citenamefont {Diddams}(2001)}]{Diddams2001}%
  \BibitemOpen
  \bibfield  {author} {\bibinfo {author} {\bibfnamefont {S.~A.}\ \bibnamefont
  {Diddams}},\ }\bibfield  {title} {\enquote {\bibinfo {title} {An optical
  clock based on a single trapped $^{199}$ {Hg} $^+$ ion},}\ }\href {\doibase
  10.1126/science.1061171} {\bibfield  {journal} {\bibinfo  {journal}
  {Science}\ }\textbf {\bibinfo {volume} {293}},\ \bibinfo {pages} {825--828}
  (\bibinfo {year} {2001})}\BibitemShut {NoStop}%
\bibitem [{\citenamefont {Washburn}\ \emph {et~al.}(2004)\citenamefont
  {Washburn}, \citenamefont {Diddams}, \citenamefont {Newbury}, \citenamefont
  {Nicholson}, \citenamefont {Yan},\ and\ \citenamefont
  {J{\o}rgensen}}]{Washburn2004}%
  \BibitemOpen
  \bibfield  {author} {\bibinfo {author} {\bibfnamefont {B.~R.}\ \bibnamefont
  {Washburn}}, \bibinfo {author} {\bibfnamefont {S.~A.}\ \bibnamefont
  {Diddams}}, \bibinfo {author} {\bibfnamefont {N.~R.}\ \bibnamefont
  {Newbury}}, \bibinfo {author} {\bibfnamefont {J.~W.}\ \bibnamefont
  {Nicholson}}, \bibinfo {author} {\bibfnamefont {M.~F.}\ \bibnamefont {Yan}},
  \ and\ \bibinfo {author} {\bibfnamefont {C.~G.}\ \bibnamefont
  {J{\o}rgensen}},\ }\bibfield  {title} {\enquote {\bibinfo {title}
  {Phase-locked, erbium-fiber-laser-based frequency comb in the near
  infrared},}\ }\href {\doibase 10.1364/OL.29.000250} {\bibfield  {journal}
  {\bibinfo  {journal} {Opt. Lett.}\ }\textbf {\bibinfo {volume} {29}},\
  \bibinfo {pages} {250} (\bibinfo {year} {2004})}\BibitemShut {NoStop}%
\bibitem [{\citenamefont {Diddams}, \citenamefont {Hollberg},\ and\
  \citenamefont {Mbele}(2007)}]{Diddams2007}%
  \BibitemOpen
  \bibfield  {author} {\bibinfo {author} {\bibfnamefont {S.~A.}\ \bibnamefont
  {Diddams}}, \bibinfo {author} {\bibfnamefont {L.}~\bibnamefont {Hollberg}}, \
  and\ \bibinfo {author} {\bibfnamefont {V.}~\bibnamefont {Mbele}},\ }\bibfield
   {title} {\enquote {\bibinfo {title} {Molecular fingerprinting with the
  resolved modes of a femtosecond laser frequency comb},}\ }\href {\doibase
  10.1038/nature05524} {\bibfield  {journal} {\bibinfo  {journal} {Nature}\
  }\textbf {\bibinfo {volume} {445}},\ \bibinfo {pages} {627--630} (\bibinfo
  {year} {2007})}\BibitemShut {NoStop}%
\bibitem [{\citenamefont {{McRaven}}\ \emph {et~al.}(2011)\citenamefont
  {{McRaven}}, \citenamefont {Cich}, \citenamefont {Lopez}, \citenamefont
  {Sears}, \citenamefont {Hurtmans},\ and\ \citenamefont
  {Mantz}}]{Mcraven_2011}%
  \BibitemOpen
  \bibfield  {author} {\bibinfo {author} {\bibfnamefont {C.}~\bibnamefont
  {{McRaven}}}, \bibinfo {author} {\bibfnamefont {M.}~\bibnamefont {Cich}},
  \bibinfo {author} {\bibfnamefont {G.}~\bibnamefont {Lopez}}, \bibinfo
  {author} {\bibfnamefont {T.~J.}\ \bibnamefont {Sears}}, \bibinfo {author}
  {\bibfnamefont {D.}~\bibnamefont {Hurtmans}}, \ and\ \bibinfo {author}
  {\bibfnamefont {A.}~\bibnamefont {Mantz}},\ }\bibfield  {title} {\enquote
  {\bibinfo {title} {Frequency comb-referenced measurements of self- and
  nitrogen-broadening in the $\nu_{1}+\nu_{3}$ band of acetylene},}\ }\href
  {\doibase 10.1016/j.jms.2011.02.016} {\bibfield  {journal} {\bibinfo
  {journal} {J. Mol. Spectrosc.}\ }\textbf {\bibinfo {volume} {266}},\ \bibinfo
  {pages} {43--51} (\bibinfo {year} {2011})}\BibitemShut {NoStop}%
\bibitem [{\citenamefont {Anderson}(1949)}]{Anderson1949}%
  \BibitemOpen
  \bibfield  {author} {\bibinfo {author} {\bibfnamefont {P.~W.}\ \bibnamefont
  {Anderson}},\ }\bibfield  {title} {\enquote {\bibinfo {title} {Pressure
  broadening in the microwave and infra-red regions},}\ }\href {\doibase
  10.1103/PhysRev.76.647} {\bibfield  {journal} {\bibinfo  {journal} {Phys.
  Rev.}\ }\textbf {\bibinfo {volume} {76}},\ \bibinfo {pages} {647--661}
  (\bibinfo {year} {1949})}\BibitemShut {NoStop}%
\bibitem [{\citenamefont {Keijser}\ \emph {et~al.}(1974)\citenamefont
  {Keijser}, \citenamefont {Lombardi}, \citenamefont {den Hout}, \citenamefont
  {Sanctuary},\ and\ \citenamefont {Knaap}}]{Keijser1974}%
  \BibitemOpen
  \bibfield  {author} {\bibinfo {author} {\bibfnamefont {R.}~\bibnamefont
  {Keijser}}, \bibinfo {author} {\bibfnamefont {J.}~\bibnamefont {Lombardi}},
  \bibinfo {author} {\bibfnamefont {K.~V.}\ \bibnamefont {den Hout}}, \bibinfo
  {author} {\bibfnamefont {B.}~\bibnamefont {Sanctuary}}, \ and\ \bibinfo
  {author} {\bibfnamefont {H.}~\bibnamefont {Knaap}},\ }\bibfield  {title}
  {\enquote {\bibinfo {title} {The pressure broadening of the rotational
  {R}aman lines of hydrogen isotopes},}\ }\href {\doibase
  https://doi.org/10.1016/0031-8914(74)90160-8} {\bibfield  {journal} {\bibinfo
   {journal} {Physica}\ }\textbf {\bibinfo {volume} {76}},\ \bibinfo {pages}
  {585 -- 608} (\bibinfo {year} {1974})}\BibitemShut {NoStop}%
\bibitem [{\citenamefont {Hout}\ \emph {et~al.}(1980)\citenamefont {Hout},
  \citenamefont {Hermans}, \citenamefont {Mazur},\ and\ \citenamefont
  {Knaap}}]{VanDenHout1988}%
  \BibitemOpen
  \bibfield  {author} {\bibinfo {author} {\bibfnamefont {K.~V.~D.}\
  \bibnamefont {Hout}}, \bibinfo {author} {\bibfnamefont {P.}~\bibnamefont
  {Hermans}}, \bibinfo {author} {\bibfnamefont {E.}~\bibnamefont {Mazur}}, \
  and\ \bibinfo {author} {\bibfnamefont {H.}~\bibnamefont {Knaap}},\ }\bibfield
   {title} {\enquote {\bibinfo {title} {The broadening and shift of the
  ratational raman lines for hydrogen isotopes at low temperatures},}\ }\href
  {\doibase https://doi.org/10.1016/0378-4371(80)90012-6} {\bibfield  {journal}
  {\bibinfo  {journal} {Physica A: Statistical Mechanics and its Applications}\
  }\textbf {\bibinfo {volume} {104}},\ \bibinfo {pages} {509 -- 547} (\bibinfo
  {year} {1980})}\BibitemShut {NoStop}%
\bibitem [{\citenamefont {Rahn}, \citenamefont {Farrow},\ and\ \citenamefont
  {Rosasco}(1991)}]{Rahn1991}%
  \BibitemOpen
  \bibfield  {author} {\bibinfo {author} {\bibfnamefont {L.~A.}\ \bibnamefont
  {Rahn}}, \bibinfo {author} {\bibfnamefont {R.~L.}\ \bibnamefont {Farrow}}, \
  and\ \bibinfo {author} {\bibfnamefont {G.~J.}\ \bibnamefont {Rosasco}},\
  }\bibfield  {title} {\enquote {\bibinfo {title} {Measurement of the
  self-broadening of the {H}$_{2}$ {Q}(0--5) {Raman} transitions from 295 to
  1000 {K}},}\ }\href {\doibase 10.1103/PhysRevA.43.6075} {\bibfield  {journal}
  {\bibinfo  {journal} {Phys. Rev. A}\ }\textbf {\bibinfo {volume} {43}},\
  \bibinfo {pages} {6075--6088} (\bibinfo {year} {1991})}\BibitemShut {NoStop}%
\bibitem [{\citenamefont {Gray}(1971)}]{Gray1971}%
  \BibitemOpen
  \bibfield  {author} {\bibinfo {author} {\bibfnamefont {C.}~\bibnamefont
  {Gray}},\ }\bibfield  {title} {\enquote {\bibinfo {title}
  {Pressure-broadening of the rotational {R}aman lines of {HCl}},}\ }\href
  {\doibase https://doi.org/10.1016/0009-2614(71)80083-0} {\bibfield  {journal}
  {\bibinfo  {journal} {Chem. Phys. Lett.}\ }\textbf {\bibinfo {volume} {8}},\
  \bibinfo {pages} {527 -- 528} (\bibinfo {year} {1971})}\BibitemShut {NoStop}%
\bibitem [{\citenamefont {Rich}\ and\ \citenamefont {Welsh}(1971)}]{Rich1971}%
  \BibitemOpen
  \bibfield  {author} {\bibinfo {author} {\bibfnamefont {N.}~\bibnamefont
  {Rich}}\ and\ \bibinfo {author} {\bibfnamefont {H.}~\bibnamefont {Welsh}},\
  }\bibfield  {title} {\enquote {\bibinfo {title} {Measurement of the pressure
  broadening of the rotational {R}aman lines of {HCl}},}\ }\href {\doibase
  https://doi.org/10.1016/0009-2614(71)80488-8} {\bibfield  {journal} {\bibinfo
   {journal} {Chem. Phys. Lett.}\ }\textbf {\bibinfo {volume} {11}},\ \bibinfo
  {pages} {292 -- 293} (\bibinfo {year} {1971})}\BibitemShut {NoStop}%
\bibitem [{\citenamefont {Fabre}, \citenamefont {Widenlocher},\ and\
  \citenamefont {Vu}(1972)}]{Fabre1972}%
  \BibitemOpen
  \bibfield  {author} {\bibinfo {author} {\bibfnamefont {D.}~\bibnamefont
  {Fabre}}, \bibinfo {author} {\bibfnamefont {G.}~\bibnamefont {Widenlocher}},
  \ and\ \bibinfo {author} {\bibfnamefont {H.}~\bibnamefont {Vu}},\ }\bibfield
  {title} {\enquote {\bibinfo {title} {Etude experimentale de l'elargissement
  par la pression des raies de rotation du spectre raman de l'acide
  chlorhydrique},}\ }\href {\doibase
  https://doi.org/10.1016/0030-4018(72)90115-0} {\bibfield  {journal} {\bibinfo
   {journal} {Opt. Commun.}\ }\textbf {\bibinfo {volume} {4}},\ \bibinfo
  {pages} {421 -- 424} (\bibinfo {year} {1972})}\BibitemShut {NoStop}%
\bibitem [{\citenamefont {Smith}, \citenamefont {Lehmann},\ and\ \citenamefont
  {Klemperer}(1986)}]{Smith1986}%
  \BibitemOpen
  \bibfield  {author} {\bibinfo {author} {\bibfnamefont {A.~M.}\ \bibnamefont
  {Smith}}, \bibinfo {author} {\bibfnamefont {K.~K.}\ \bibnamefont {Lehmann}},
  \ and\ \bibinfo {author} {\bibfnamefont {W.}~\bibnamefont {Klemperer}},\
  }\bibfield  {title} {\enquote {\bibinfo {title} {The intensity and
  self‐broadening of overtone transitions in {HCN}},}\ }\href {\doibase
  10.1063/1.451734} {\bibfield  {journal} {\bibinfo  {journal} {J. Chem.
  Phys.}\ }\textbf {\bibinfo {volume} {85}},\ \bibinfo {pages} {4958--4965}
  (\bibinfo {year} {1986})},\ \Eprint
  {http://arxiv.org/abs/https://doi.org/10.1063/1.451734}
  {https://doi.org/10.1063/1.451734} \BibitemShut {NoStop}%
\bibitem [{\citenamefont {Swann}\ and\ \citenamefont
  {Gilbert}(2005)}]{Swann2005}%
  \BibitemOpen
  \bibfield  {author} {\bibinfo {author} {\bibfnamefont {W.}~\bibnamefont
  {Swann}}\ and\ \bibinfo {author} {\bibfnamefont {S.}~\bibnamefont
  {Gilbert}},\ }\bibfield  {title} {\enquote {\bibinfo {title} {{Line centers,
  pressure shift, and pressure broadening of 1530-1560 nm hydrogen cyanide
  wavelength calibration lines}},}\ }\href {\doibase {10.1364/JOSAB.22.001749}}
  {\bibfield  {journal} {\bibinfo  {journal} {{J. Opt. Soc. Am. B-Opt.Phys.}}\
  }\textbf {\bibinfo {volume} {{22}}},\ \bibinfo {pages} {{1749--1756}}
  (\bibinfo {year} {{2005}})}\BibitemShut {NoStop}%
\bibitem [{\citenamefont {Bouanich}\ \emph {et~al.}(2005)\citenamefont
  {Bouanich}, \citenamefont {Boulet}, \citenamefont {Predoi-Cross},
  \citenamefont {Sharpe}, \citenamefont {Sams}, \citenamefont {Smith},
  \citenamefont {Rinsland}, \citenamefont {Benner},\ and\ \citenamefont
  {Devi}}]{Bouanich2005}%
  \BibitemOpen
  \bibfield  {author} {\bibinfo {author} {\bibfnamefont {J.}~\bibnamefont
  {Bouanich}}, \bibinfo {author} {\bibfnamefont {C.}~\bibnamefont {Boulet}},
  \bibinfo {author} {\bibfnamefont {A.}~\bibnamefont {Predoi-Cross}}, \bibinfo
  {author} {\bibfnamefont {S.}~\bibnamefont {Sharpe}}, \bibinfo {author}
  {\bibfnamefont {R.}~\bibnamefont {Sams}}, \bibinfo {author} {\bibfnamefont
  {M.}~\bibnamefont {Smith}}, \bibinfo {author} {\bibfnamefont
  {C.}~\bibnamefont {Rinsland}}, \bibinfo {author} {\bibfnamefont
  {D.}~\bibnamefont {Benner}}, \ and\ \bibinfo {author} {\bibfnamefont
  {V.}~\bibnamefont {Devi}},\ }\bibfield  {title} {\enquote {\bibinfo {title}
  {{A multispectrum analysis of the $\nu_2$ band of {(HCN)-C-12-N-14}: Part II.
  Theoretical calculations of self-broadening, self-induced shifts, and their
  temperature dependences}},}\ }\href {\doibase {10.1016/j.jms.2004.12.002}}
  {\bibfield  {journal} {\bibinfo  {journal} {{J. Mol. Spectrosc.}}\ }\textbf
  {\bibinfo {volume} {{231}}},\ \bibinfo {pages} {{85--95}} (\bibinfo {year}
  {{2005}})}\BibitemShut {NoStop}%
\bibitem [{\citenamefont {Lehmann}(2017)}]{Lehmann2017}%
  \BibitemOpen
  \bibfield  {author} {\bibinfo {author} {\bibfnamefont {K.~K.}\ \bibnamefont
  {Lehmann}},\ }\bibfield  {title} {\enquote {\bibinfo {title} {{Influence of
  resonant collisions on the self-broadening of acetylene}},}\ }\href {\doibase
  {10.1063/1.4977726}} {\bibfield  {journal} {\bibinfo  {journal} {{J.Chem.
  Phys.}}\ }\textbf {\bibinfo {volume} {{146}}},\ \bibinfo {pages} {094309}
  (\bibinfo {year} {{2017}})}\BibitemShut {NoStop}%
\bibitem [{\citenamefont {Hartmann}\ and\ \citenamefont
  {Tran}(2017)}]{Hartmann2017}%
  \BibitemOpen
  \bibfield  {author} {\bibinfo {author} {\bibfnamefont {J.-M.}\ \bibnamefont
  {Hartmann}}\ and\ \bibinfo {author} {\bibfnamefont {H.}~\bibnamefont
  {Tran}},\ }\bibfield  {title} {\enquote {\bibinfo {title} {Comment on
  ``ortho-para-dependent pressure effects observed in the near infrared band of
  acetylene by dual-comb spectroscopy''},}\ }\href {\doibase
  10.1103/PhysRevLett.119.069401} {\bibfield  {journal} {\bibinfo  {journal}
  {Phys. Rev. Lett.}\ }\textbf {\bibinfo {volume} {119}},\ \bibinfo {pages}
  {069401} (\bibinfo {year} {2017})}\BibitemShut {NoStop}%
\bibitem [{\citenamefont {Iwakuni}\ \emph {et~al.}(2017)\citenamefont
  {Iwakuni}, \citenamefont {Okubo}, \citenamefont {Yamada}, \citenamefont
  {Inaba}, \citenamefont {Onae}, \citenamefont {Hong},\ and\ \citenamefont
  {Sasada}}]{Iwakuni2017}%
  \BibitemOpen
  \bibfield  {author} {\bibinfo {author} {\bibfnamefont {K.}~\bibnamefont
  {Iwakuni}}, \bibinfo {author} {\bibfnamefont {S.}~\bibnamefont {Okubo}},
  \bibinfo {author} {\bibfnamefont {K.~M.~T.}\ \bibnamefont {Yamada}}, \bibinfo
  {author} {\bibfnamefont {H.}~\bibnamefont {Inaba}}, \bibinfo {author}
  {\bibfnamefont {A.}~\bibnamefont {Onae}}, \bibinfo {author} {\bibfnamefont
  {F.-L.}\ \bibnamefont {Hong}}, \ and\ \bibinfo {author} {\bibfnamefont
  {H.}~\bibnamefont {Sasada}},\ }\bibfield  {title} {\enquote {\bibinfo {title}
  {Iwakuni et al. reply:},}\ }\href {\doibase 10.1103/PhysRevLett.119.069402}
  {\bibfield  {journal} {\bibinfo  {journal} {Phys. Rev. Lett.}\ }\textbf
  {\bibinfo {volume} {119}},\ \bibinfo {pages} {069402} (\bibinfo {year}
  {2017})}\BibitemShut {NoStop}%
\bibitem [{\citenamefont {Ngo}, \citenamefont {Tran},\ and\ \citenamefont
  {Hartmann}(2013)}]{Ngo2013}%
  \BibitemOpen
  \bibfield  {author} {\bibinfo {author} {\bibfnamefont {N.~H.}\ \bibnamefont
  {Ngo}}, \bibinfo {author} {\bibfnamefont {H.}~\bibnamefont {Tran}}, \ and\
  \bibinfo {author} {\bibfnamefont {J.~M.}\ \bibnamefont {Hartmann}},\
  }\bibfield  {title} {\enquote {\bibinfo {title} {An isolated line-shape model
  to go beyond the {V}oigt profile in spectroscopic databases and radiative
  transfer codes},}\ }\href@noop {} {\bibfield  {journal} {\bibinfo  {journal}
  {J. Quant. Spectros. Radiat. Transfer}\ }\textbf {\bibinfo {volume} {129}},\
  \bibinfo {pages} {89--100} (\bibinfo {year} {2013})}\BibitemShut {NoStop}%
\bibitem [{\citenamefont {Tennyson}\ \emph {et~al.}(2014)\citenamefont
  {Tennyson}, \citenamefont {Bernath}, \citenamefont {Campargue}, \citenamefont
  {Császár}, \citenamefont {Daumont}, \citenamefont {Gamache}, \citenamefont
  {Hodges}, \citenamefont {Lisak}, \citenamefont {Naumenko}, \citenamefont
  {Rothman}, \citenamefont {Tran}, \citenamefont {Zobov}, \citenamefont
  {Buldyreva}, \citenamefont {Boone}, \citenamefont {De~Vizia}, \citenamefont
  {Gianfrani}, \citenamefont {Hartmann}, \citenamefont {McPheat}, \citenamefont
  {Weidmann}, \citenamefont {Murray}, \citenamefont {Ngo},\ and\ \citenamefont
  {Polyansky}}]{Tennyson2014_IUPAC}%
  \BibitemOpen
  \bibfield  {author} {\bibinfo {author} {\bibfnamefont {J.}~\bibnamefont
  {Tennyson}}, \bibinfo {author} {\bibfnamefont {P.~F.}\ \bibnamefont
  {Bernath}}, \bibinfo {author} {\bibfnamefont {A.}~\bibnamefont {Campargue}},
  \bibinfo {author} {\bibfnamefont {A.~G.}\ \bibnamefont {Császár}}, \bibinfo
  {author} {\bibfnamefont {L.}~\bibnamefont {Daumont}}, \bibinfo {author}
  {\bibfnamefont {R.~R.}\ \bibnamefont {Gamache}}, \bibinfo {author}
  {\bibfnamefont {J.~T.}\ \bibnamefont {Hodges}}, \bibinfo {author}
  {\bibfnamefont {D.}~\bibnamefont {Lisak}}, \bibinfo {author} {\bibfnamefont
  {O.~V.}\ \bibnamefont {Naumenko}}, \bibinfo {author} {\bibfnamefont {L.~S.}\
  \bibnamefont {Rothman}}, \bibinfo {author} {\bibfnamefont {H.}~\bibnamefont
  {Tran}}, \bibinfo {author} {\bibfnamefont {N.~F.}\ \bibnamefont {Zobov}},
  \bibinfo {author} {\bibfnamefont {J.}~\bibnamefont {Buldyreva}}, \bibinfo
  {author} {\bibfnamefont {C.~D.}\ \bibnamefont {Boone}}, \bibinfo {author}
  {\bibfnamefont {M.~D.}\ \bibnamefont {De~Vizia}}, \bibinfo {author}
  {\bibfnamefont {L.}~\bibnamefont {Gianfrani}}, \bibinfo {author}
  {\bibfnamefont {J.-M.}\ \bibnamefont {Hartmann}}, \bibinfo {author}
  {\bibfnamefont {R.}~\bibnamefont {McPheat}}, \bibinfo {author} {\bibfnamefont
  {D.}~\bibnamefont {Weidmann}}, \bibinfo {author} {\bibfnamefont
  {J.}~\bibnamefont {Murray}}, \bibinfo {author} {\bibfnamefont {N.~H.}\
  \bibnamefont {Ngo}}, \ and\ \bibinfo {author} {\bibfnamefont {O.~L.}\
  \bibnamefont {Polyansky}},\ }\bibfield  {title} {\enquote {\bibinfo {title}
  {Recommended isolated-line profile for representing high-resolution
  spectroscopic transitions ({IUPAC} {Technical} {Report})},}\ }\href {\doibase
  10.1515/pac-2014-0208} {\bibfield  {journal} {\bibinfo  {journal} {Pure Appl.
  Chem.}\ }\textbf {\bibinfo {volume} {86}},\ \bibinfo {pages} {1931--1943}
  (\bibinfo {year} {2014})}\BibitemShut {NoStop}%
\bibitem [{\citenamefont {Adkins}\ and\ \citenamefont
  {Hodges}(2019)}]{Adkins2019}%
  \BibitemOpen
  \bibfield  {author} {\bibinfo {author} {\bibfnamefont {E.}~\bibnamefont
  {Adkins}}\ and\ \bibinfo {author} {\bibfnamefont {J.}~\bibnamefont
  {Hodges}},\ }\bibfield  {title} {\enquote {\bibinfo {title} {Numerical
  evaluation of {H}artmann-{T}ran line profile use in synthetic, noisy
  spectra},}\ }in\ \href {\doibase 10.15278/isms.2019.FE05} {\emph {\bibinfo
  {booktitle} {Proceedings of the 74th {International} {Symposium} on
  {Molecular} {Spectroscopy}}}}\ (\bibinfo  {publisher} {University of Illinois
  at Urbana-Champaign},\ \bibinfo {address} {Urbana, Illinois USA},\ \bibinfo
  {year} {2019})\ pp.\ \bibinfo {pages} {1--1}\BibitemShut {NoStop}%
\bibitem [{\citenamefont {Cich}\ \emph {et~al.}(2013)\citenamefont {Cich},
  \citenamefont {Forthomme}, \citenamefont {McRaven}, \citenamefont {Lopez},
  \citenamefont {Hall}, \citenamefont {Sears},\ and\ \citenamefont
  {Mantz}}]{Cich2013}%
  \BibitemOpen
  \bibfield  {author} {\bibinfo {author} {\bibfnamefont {M.~J.}\ \bibnamefont
  {Cich}}, \bibinfo {author} {\bibfnamefont {D.}~\bibnamefont {Forthomme}},
  \bibinfo {author} {\bibfnamefont {C.~P.}\ \bibnamefont {McRaven}}, \bibinfo
  {author} {\bibfnamefont {G.~V.}\ \bibnamefont {Lopez}}, \bibinfo {author}
  {\bibfnamefont {G.~E.}\ \bibnamefont {Hall}}, \bibinfo {author}
  {\bibfnamefont {T.~J.}\ \bibnamefont {Sears}}, \ and\ \bibinfo {author}
  {\bibfnamefont {A.~W.}\ \bibnamefont {Mantz}},\ }\bibfield  {title} {\enquote
  {\bibinfo {title} {Temperature-dependent, nitrogen-perturbed line shape
  measurements in the v$_1$ + v$_3$ band of acetylene using a diode laser
  referenced to a frequency comb},}\ }\href {\doibase 10.1021/jp408960e}
  {\bibfield  {journal} {\bibinfo  {journal} {J. Phys. Chem. A}\ }\textbf
  {\bibinfo {volume} {117}},\ \bibinfo {pages} {13908--13918} (\bibinfo {year}
  {2013})},\ \Eprint {http://arxiv.org/abs/http://dx.doi.org/10.1021/jp408960e}
  {http://dx.doi.org/10.1021/jp408960e} \BibitemShut {NoStop}%
\bibitem [{\citenamefont {McRaven}\ \emph {et~al.}(2011)\citenamefont
  {McRaven}, \citenamefont {Cich}, \citenamefont {Lopez}, \citenamefont
  {Sears}, \citenamefont {Hurtmans},\ and\ \citenamefont
  {Mantz}}]{McRaven2011}%
  \BibitemOpen
  \bibfield  {author} {\bibinfo {author} {\bibfnamefont {C.}~\bibnamefont
  {McRaven}}, \bibinfo {author} {\bibfnamefont {M.}~\bibnamefont {Cich}},
  \bibinfo {author} {\bibfnamefont {G.}~\bibnamefont {Lopez}}, \bibinfo
  {author} {\bibfnamefont {T.~J.}\ \bibnamefont {Sears}}, \bibinfo {author}
  {\bibfnamefont {D.}~\bibnamefont {Hurtmans}}, \ and\ \bibinfo {author}
  {\bibfnamefont {A.}~\bibnamefont {Mantz}},\ }\bibfield  {title} {\enquote
  {\bibinfo {title} {Frequency comb-referenced measurements of self- and
  nitrogen-broadening in the $\nu_1+\nu_3$ band of acetylene},}\ }\href
  {\doibase https://doi.org/10.1016/j.jms.2011.02.016} {\bibfield  {journal}
  {\bibinfo  {journal} {J. Mol. Spectrosc.}\ }\textbf {\bibinfo {volume}
  {266}},\ \bibinfo {pages} {43 -- 51} (\bibinfo {year} {2011})}\BibitemShut
  {NoStop}%
\bibitem [{\citenamefont {{\v{S}}ime{\v{c}}kov{\'a}}\ \emph
  {et~al.}(2006)\citenamefont {{\v{S}}ime{\v{c}}kov{\'a}}, \citenamefont
  {Jacquemart}, \citenamefont {Rothman}, \citenamefont {Gamache},\ and\
  \citenamefont {Goldman}}]{Simeckova2006}%
  \BibitemOpen
  \bibfield  {author} {\bibinfo {author} {\bibfnamefont {M.}~\bibnamefont
  {{\v{S}}ime{\v{c}}kov{\'a}}}, \bibinfo {author} {\bibfnamefont
  {D.}~\bibnamefont {Jacquemart}}, \bibinfo {author} {\bibfnamefont {L.~S.}\
  \bibnamefont {Rothman}}, \bibinfo {author} {\bibfnamefont {R.~R.}\
  \bibnamefont {Gamache}}, \ and\ \bibinfo {author} {\bibfnamefont
  {A.}~\bibnamefont {Goldman}},\ }\bibfield  {title} {\enquote {\bibinfo
  {title} {Einstein {A}-coefficients and statistical weights for molecular
  absorption transitions in the {HITRAN} database},}\ }\href@noop {} {\bibfield
   {journal} {\bibinfo  {journal} {Journal of Quant. Spec. Radiat. Transf.}\
  }\textbf {\bibinfo {volume} {98}},\ \bibinfo {pages} {130--155} (\bibinfo
  {year} {2006})}\BibitemShut {NoStop}%
\bibitem [{\citenamefont {Gamache}\ \emph {et~al.}(2000)\citenamefont
  {Gamache}, \citenamefont {Kennedy}, \citenamefont {Hawkins},\ and\
  \citenamefont {Rothman}}]{Gamache2000}%
  \BibitemOpen
  \bibfield  {author} {\bibinfo {author} {\bibfnamefont {R.}~\bibnamefont
  {Gamache}}, \bibinfo {author} {\bibfnamefont {S.}~\bibnamefont {Kennedy}},
  \bibinfo {author} {\bibfnamefont {R.}~\bibnamefont {Hawkins}}, \ and\
  \bibinfo {author} {\bibfnamefont {L.}~\bibnamefont {Rothman}},\ }\bibfield
  {title} {\enquote {\bibinfo {title} {Total internal partition sums for
  molecules in the terrestrial atmosphere},}\ }\href@noop {} {\bibfield
  {journal} {\bibinfo  {journal} {J. Mol. Struct.}\ }\textbf {\bibinfo {volume}
  {517}},\ \bibinfo {pages} {407--425} (\bibinfo {year} {2000})}\BibitemShut
  {NoStop}%
\bibitem [{\citenamefont {Amyay}, \citenamefont {Fayt},\ and\ \citenamefont
  {Herman}(2011)}]{Amyay2011}%
  \BibitemOpen
  \bibfield  {author} {\bibinfo {author} {\bibfnamefont {B.}~\bibnamefont
  {Amyay}}, \bibinfo {author} {\bibfnamefont {A.}~\bibnamefont {Fayt}}, \ and\
  \bibinfo {author} {\bibfnamefont {M.}~\bibnamefont {Herman}},\ }\bibfield
  {title} {\enquote {\bibinfo {title} {Accurate partition function for
  acetylene, $^{12}${C}$_2${H}$_2$, and related thermodynamical quantities},}\
  }\href {\doibase 10.1063/1.3664626} {\bibfield  {journal} {\bibinfo
  {journal} {J. Chem. Phys.}\ }\textbf {\bibinfo {volume} {135}},\ \bibinfo
  {pages} {234305} (\bibinfo {year} {2011})},\ \Eprint
  {http://arxiv.org/abs/https://doi.org/10.1063/1.3664626}
  {https://doi.org/10.1063/1.3664626} \BibitemShut {NoStop}%
\bibitem [{\citenamefont {Robert}\ \emph {et~al.}(2007)\citenamefont {Robert},
  \citenamefont {Herman}, \citenamefont {Auwera}, \citenamefont {Lonardo},
  \citenamefont {Fusina}, \citenamefont {Blanquet}, \citenamefont {Lepere},\
  and\ \citenamefont {Fayt}}]{Robert2007}%
  \BibitemOpen
  \bibfield  {author} {\bibinfo {author} {\bibfnamefont {S.}~\bibnamefont
  {Robert}}, \bibinfo {author} {\bibfnamefont {M.}~\bibnamefont {Herman}},
  \bibinfo {author} {\bibfnamefont {J.~V.}\ \bibnamefont {Auwera}}, \bibinfo
  {author} {\bibfnamefont {G.~D.}\ \bibnamefont {Lonardo}}, \bibinfo {author}
  {\bibfnamefont {L.}~\bibnamefont {Fusina}}, \bibinfo {author} {\bibfnamefont
  {G.}~\bibnamefont {Blanquet}}, \bibinfo {author} {\bibfnamefont
  {M.}~\bibnamefont {Lepere}}, \ and\ \bibinfo {author} {\bibfnamefont
  {A.}~\bibnamefont {Fayt}},\ }\bibfield  {title} {\enquote {\bibinfo {title}
  {The bending vibrations in $^{12}${C}$_2${H}$_2$: global vibration–rotation
  analysis},}\ }\href {\doibase 10.1080/00268970601099261} {\bibfield
  {journal} {\bibinfo  {journal} {Mol. Phys.}\ }\textbf {\bibinfo {volume}
  {105}},\ \bibinfo {pages} {559--568} (\bibinfo {year} {2007})},\ \Eprint
  {http://arxiv.org/abs/https://doi.org/10.1080/00268970601099261}
  {https://doi.org/10.1080/00268970601099261} \BibitemShut {NoStop}%
\bibitem [{\citenamefont {Gamache}\ \emph {et~al.}(2017)\citenamefont
  {Gamache}, \citenamefont {Roller}, \citenamefont {Lopes}, \citenamefont
  {Gordon}, \citenamefont {Rothman}, \citenamefont {Polyansky}, \citenamefont
  {Zobov}, \citenamefont {Kyuberis}, \citenamefont {Tennyson}, \citenamefont
  {Yurchenko}, \citenamefont {Császár}, \citenamefont {Furtenbacher},
  \citenamefont {Huang}, \citenamefont {Schwenke}, \citenamefont {Lee},
  \citenamefont {Drouin}, \citenamefont {Tashkun}, \citenamefont {Perevalov},\
  and\ \citenamefont {Kochanov}}]{Gamache2017}%
  \BibitemOpen
  \bibfield  {author} {\bibinfo {author} {\bibfnamefont {R.~R.}\ \bibnamefont
  {Gamache}}, \bibinfo {author} {\bibfnamefont {C.}~\bibnamefont {Roller}},
  \bibinfo {author} {\bibfnamefont {E.}~\bibnamefont {Lopes}}, \bibinfo
  {author} {\bibfnamefont {I.~E.}\ \bibnamefont {Gordon}}, \bibinfo {author}
  {\bibfnamefont {L.~S.}\ \bibnamefont {Rothman}}, \bibinfo {author}
  {\bibfnamefont {O.~L.}\ \bibnamefont {Polyansky}}, \bibinfo {author}
  {\bibfnamefont {N.~F.}\ \bibnamefont {Zobov}}, \bibinfo {author}
  {\bibfnamefont {A.~A.}\ \bibnamefont {Kyuberis}}, \bibinfo {author}
  {\bibfnamefont {J.}~\bibnamefont {Tennyson}}, \bibinfo {author}
  {\bibfnamefont {S.~N.}\ \bibnamefont {Yurchenko}}, \bibinfo {author}
  {\bibfnamefont {A.~G.}\ \bibnamefont {Császár}}, \bibinfo {author}
  {\bibfnamefont {T.}~\bibnamefont {Furtenbacher}}, \bibinfo {author}
  {\bibfnamefont {X.}~\bibnamefont {Huang}}, \bibinfo {author} {\bibfnamefont
  {D.~W.}\ \bibnamefont {Schwenke}}, \bibinfo {author} {\bibfnamefont {T.~J.}\
  \bibnamefont {Lee}}, \bibinfo {author} {\bibfnamefont {B.~J.}\ \bibnamefont
  {Drouin}}, \bibinfo {author} {\bibfnamefont {S.~A.}\ \bibnamefont {Tashkun}},
  \bibinfo {author} {\bibfnamefont {V.~I.}\ \bibnamefont {Perevalov}}, \ and\
  \bibinfo {author} {\bibfnamefont {R.~V.}\ \bibnamefont {Kochanov}},\
  }\bibfield  {title} {\enquote {\bibinfo {title} {{Total internal partition
  sums for 166 isotopologues of 51 molecules important in planetary atmospheres
  Application to {HITRAN2016} and beyond}},}\ }\href {\doibase
  https://doi.org/10.1016/j.jqsrt.2017.03.045} {\bibfield  {journal} {\bibinfo
  {journal} {J. {Q}uant. {S}pectros. {R}adiat. {T}ransfer}\ }\textbf {\bibinfo
  {volume} {203}},\ \bibinfo {pages} {70 -- 87} (\bibinfo {year} {2017})},\
  \bibinfo {note} {hITRAN2016 Special Issue}\BibitemShut {NoStop}%
\bibitem [{\citenamefont {Townes}\ and\ \citenamefont
  {Schawlow}(1975)}]{Townes1975}%
  \BibitemOpen
  \bibfield  {author} {\bibinfo {author} {\bibfnamefont {C.~H.}\ \bibnamefont
  {Townes}}\ and\ \bibinfo {author} {\bibfnamefont {A.~L.}\ \bibnamefont
  {Schawlow}},\ }\href@noop {} {\emph {\bibinfo {title} {Microwave
  {S}pectroscopy}}}\ (\bibinfo  {publisher} {Dover Publications},\ \bibinfo
  {address} {New York},\ \bibinfo {year} {1975})\BibitemShut {NoStop}%
\bibitem [{\citenamefont {Forthomme}\ \emph {et~al.}(2015)\citenamefont
  {Forthomme}, \citenamefont {Cich}, \citenamefont {Twagirayezu}, \citenamefont
  {Hall},\ and\ \citenamefont {Sears}}]{Forthomme2015}%
  \BibitemOpen
  \bibfield  {author} {\bibinfo {author} {\bibfnamefont {D.}~\bibnamefont
  {Forthomme}}, \bibinfo {author} {\bibfnamefont {M.~J.}\ \bibnamefont {Cich}},
  \bibinfo {author} {\bibfnamefont {S.}~\bibnamefont {Twagirayezu}}, \bibinfo
  {author} {\bibfnamefont {G.~E.}\ \bibnamefont {Hall}}, \ and\ \bibinfo
  {author} {\bibfnamefont {T.~J.}\ \bibnamefont {Sears}},\ }\bibfield  {title}
  {\enquote {\bibinfo {title} {{Application of the Hartmann–Tran profile to
  precise experimental data sets of $^{12}$C$_2$H$_2$}},}\ }\href@noop {}
  {\bibfield  {journal} {\bibinfo  {journal} {J. Quant. Spectros. Radiat.
  Transfer.}\ }\textbf {\bibinfo {volume} {165}},\ \bibinfo {pages} {28--37}
  (\bibinfo {year} {2015})}\BibitemShut {NoStop}%
\bibitem [{\citenamefont {Okubo}\ \emph {et~al.}(2017)\citenamefont {Okubo},
  \citenamefont {Iwakuni}, \citenamefont {Yamada}, \citenamefont {Inaba},
  \citenamefont {Onae}, \citenamefont {Hong},\ and\ \citenamefont
  {Sasada}}]{Okubo2017}%
  \BibitemOpen
  \bibfield  {author} {\bibinfo {author} {\bibfnamefont {S.}~\bibnamefont
  {Okubo}}, \bibinfo {author} {\bibfnamefont {K.}~\bibnamefont {Iwakuni}},
  \bibinfo {author} {\bibfnamefont {K.~M.}\ \bibnamefont {Yamada}}, \bibinfo
  {author} {\bibfnamefont {H.}~\bibnamefont {Inaba}}, \bibinfo {author}
  {\bibfnamefont {A.}~\bibnamefont {Onae}}, \bibinfo {author} {\bibfnamefont
  {F.-L.}\ \bibnamefont {Hong}}, \ and\ \bibinfo {author} {\bibfnamefont
  {H.}~\bibnamefont {Sasada}},\ }\bibfield  {title} {\enquote {\bibinfo {title}
  {Transition dipole-moment of the $\nu_1 + \nu_3$ band of acetylene measured
  with dual-comb {F}ourier-transform spectroscopy},}\ }\href {\doibase
  https://doi.org/10.1016/j.jms.2017.09.001} {\bibfield  {journal} {\bibinfo
  {journal} {J. Mol. Spectrosc.}\ }\textbf {\bibinfo {volume} {341}},\ \bibinfo
  {pages} {10 -- 16} (\bibinfo {year} {2017})}\BibitemShut {NoStop}%
\bibitem [{\citenamefont {Madej}\ \emph {et~al.}(2006)\citenamefont {Madej},
  \citenamefont {Alcock}, \citenamefont {Czajkowski}, \citenamefont {Bernard},\
  and\ \citenamefont {Chepurov}}]{Madej2006}%
  \BibitemOpen
  \bibfield  {author} {\bibinfo {author} {\bibfnamefont {A.~A.}\ \bibnamefont
  {Madej}}, \bibinfo {author} {\bibfnamefont {A.~J.}\ \bibnamefont {Alcock}},
  \bibinfo {author} {\bibfnamefont {A.}~\bibnamefont {Czajkowski}}, \bibinfo
  {author} {\bibfnamefont {J.~E.}\ \bibnamefont {Bernard}}, \ and\ \bibinfo
  {author} {\bibfnamefont {S.}~\bibnamefont {Chepurov}},\ }\bibfield  {title}
  {\enquote {\bibinfo {title} {{Accurate absolute reference frequencies from
  1511 to 1545 nm of the $\nu_1 + \nu_3$ band of $^{12}$C$_2$H$_2$ determined
  with laser frequency comb interval measurements}},}\ }\href@noop {}
  {\bibfield  {journal} {\bibinfo  {journal} {J. Opt. Soc. Am. B}\ }\textbf
  {\bibinfo {volume} {23}},\ \bibinfo {pages} {2200--2208} (\bibinfo {year}
  {2006})}\BibitemShut {NoStop}%
\bibitem [{\citenamefont {Kusaba}\ and\ \citenamefont
  {Henningsen}(2001)}]{Kusaba2001}%
  \BibitemOpen
  \bibfield  {author} {\bibinfo {author} {\bibfnamefont {M.}~\bibnamefont
  {Kusaba}}\ and\ \bibinfo {author} {\bibfnamefont {J.}~\bibnamefont
  {Henningsen}},\ }\bibfield  {title} {\enquote {\bibinfo {title} {The
  $\nu_1+\nu_3$ and the $\nu_1+\nu_2+\nu_4^1+\nu_5^{-1}$ combination bands of
  $^{13}${C}$_2${H}$_2$, linestrengths, broadening parameters and pressure
  shifts},}\ }\href@noop {} {\bibfield  {journal} {\bibinfo  {journal} {J. Mol.
  Spectrosc.}\ }\textbf {\bibinfo {volume} {209}},\ \bibinfo {pages} {216--227}
  (\bibinfo {year} {2001})}\BibitemShut {NoStop}%
\bibitem [{\citenamefont {Jacquemart}\ \emph {et~al.}(2003)\citenamefont
  {Jacquemart}, \citenamefont {Mandin}, \citenamefont {Dana}, \citenamefont
  {Regalia-Jarlot}, \citenamefont {Plateaux}, \citenamefont {Decatoire},\ and\
  \citenamefont {Rothman}}]{Jacquemart2003}%
  \BibitemOpen
  \bibfield  {author} {\bibinfo {author} {\bibfnamefont {D.}~\bibnamefont
  {Jacquemart}}, \bibinfo {author} {\bibfnamefont {J.-Y.}\ \bibnamefont
  {Mandin}}, \bibinfo {author} {\bibfnamefont {V.}~\bibnamefont {Dana}},
  \bibinfo {author} {\bibfnamefont {L.}~\bibnamefont {Regalia-Jarlot}},
  \bibinfo {author} {\bibfnamefont {J.~J.}\ \bibnamefont {Plateaux}}, \bibinfo
  {author} {\bibfnamefont {D.}~\bibnamefont {Decatoire}}, \ and\ \bibinfo
  {author} {\bibfnamefont {L.~S.}\ \bibnamefont {Rothman}},\ }\bibfield
  {title} {\enquote {\bibinfo {title} {The spectrum of acetylene in the
  5-$\mu$m region from new line parameter measurements},}\ }\href {\doibase
  10.1016/S0022-4073(02)00055-9} {\bibfield  {journal} {\bibinfo  {journal} {J.
  Quant. Spectros. Radiat. Transfer}\ }\textbf {\bibinfo {volume} {76}},\
  \bibinfo {pages} {237--267} (\bibinfo {year} {2003})}\BibitemShut {NoStop}%
\bibitem [{\citenamefont {{James}}\ and\ \citenamefont
  {{Roos}}(1975)}]{James1975}%
  \BibitemOpen
  \bibfield  {author} {\bibinfo {author} {\bibfnamefont {F.}~\bibnamefont
  {{James}}}\ and\ \bibinfo {author} {\bibfnamefont {M.}~\bibnamefont
  {{Roos}}},\ }\bibfield  {title} {\enquote {\bibinfo {title} {{Minuit -- a
  system for function minimization and analysis of the parameter errors and
  correlations}},}\ }\href {\doibase 10.1016/0010-4655(75)90039-9} {\bibfield
  {journal} {\bibinfo  {journal} {Comput. Phys. Commun.}\ }\textbf {\bibinfo
  {volume} {10}},\ \bibinfo {pages} {343--367} (\bibinfo {year}
  {1975})}\BibitemShut {NoStop}%
\bibitem [{\citenamefont {Dembinski}\ \emph {et~al.}(2020)\citenamefont
  {Dembinski}, \citenamefont {Ongmongkolkul}, \citenamefont {Deil},
  \citenamefont {Hurtado}, \citenamefont {Schreiner}, \citenamefont {Feickert},
  \citenamefont {Andrew}, \citenamefont {Burr}, \citenamefont {Rost},
  \citenamefont {Pearce}, \citenamefont {Geiger}, \citenamefont {Wiedemann},
  \citenamefont {Gonzalo}, \citenamefont {Gorelli},\ and\ \citenamefont
  {Zapata}}]{iminuit2020}%
  \BibitemOpen
  \bibfield  {author} {\bibinfo {author} {\bibfnamefont {H.}~\bibnamefont
  {Dembinski}}, \bibinfo {author} {\bibfnamefont {P.}~\bibnamefont
  {Ongmongkolkul}}, \bibinfo {author} {\bibfnamefont {C.}~\bibnamefont {Deil}},
  \bibinfo {author} {\bibfnamefont {D.~M.}\ \bibnamefont {Hurtado}}, \bibinfo
  {author} {\bibfnamefont {H.}~\bibnamefont {Schreiner}}, \bibinfo {author}
  {\bibfnamefont {M.}~\bibnamefont {Feickert}}, \bibinfo {author} {\bibnamefont
  {Andrew}}, \bibinfo {author} {\bibfnamefont {C.}~\bibnamefont {Burr}},
  \bibinfo {author} {\bibfnamefont {F.}~\bibnamefont {Rost}}, \bibinfo {author}
  {\bibfnamefont {A.}~\bibnamefont {Pearce}}, \bibinfo {author} {\bibfnamefont
  {L.}~\bibnamefont {Geiger}}, \bibinfo {author} {\bibfnamefont {B.~M.}\
  \bibnamefont {Wiedemann}}, \bibinfo {author} {\bibnamefont {Gonzalo}},
  \bibinfo {author} {\bibfnamefont {M.}~\bibnamefont {Gorelli}}, \ and\
  \bibinfo {author} {\bibfnamefont {O.}~\bibnamefont {Zapata}},\ }\href
  {\doibase 10.5281/ZENODO.4299243} {\enquote {\bibinfo {title}
  {scikit-hep/iminuit v2.0rc1},}\ } (\bibinfo {year} {2020})\BibitemShut
  {NoStop}%
\end{thebibliography}
%

\end{document}